\documentclass{aa}  

\usepackage{graphicx}
\usepackage{txfonts}
\usepackage{amsmath}
\usepackage[colorlinks=true, linkcolor=blue, citecolor=blue, urlcolor=blue]{hyperref} 

\begin{document}

\titlerunning{The mid-infrared integrated optics beam combiner for NOTT}
\authorrunning{Sanny et al.}

\title{Asgard/NOTT: L-band nulling interferometry at the VLTI}
\subtitle{III. The mid-infrared integrated optics beam combiner for NOTT}

\author{
    Ahmed Sanny\inst{1,2,3}
    \and Lucas Labadie\inst{1}
    \and Simon Gross\inst{2,5}
    \and Kévin Barjot\inst{1}
    \and Romain Laugier\inst{4}
    \and Germain Garreau\inst{4}
    \and Marc-Antoine Martinod\inst{4}
    \and Denis Defrère\inst{4}
    \and Michael J. Withford\inst{2}
    }

\institute{
    I. Physikalisches Institut der Universität zu Köln, Zülpicher Str. 77, 50937 Köln, Germany \\  
    \email{ahmed@ph1.uni-koeln.de}
    \and
    MQ Photonics Research Centre, School of Mathematical and Physical Sciences, Macquarie University, New South Wales 2109, Australia
    \and
    Department of Physics \& Astronomy, University of California, Los Angeles, CA 90095, USA
    \and
    Institute of Astronomy, KU Leuven, Celestijnenlaan 200D, 3001 Leuven, Belgium
    \and 
     MQ Photonics Research Centre, School of Engineering, Macquarie University, New South Wales 2109, Australia
    }

\date{Received XX Month 2025 / Accepted XX Month 20XX}  
 
  \abstract
   {The NOTT visitor instrument at the VLTI is designed to characterize hot exozodiacal dust and young Jupiter-like planets at the water snowline via L$^{\prime}$ band nulling interferometry. The beam combination will be achieved by a four-telescope integrated optics beam combiner, which should fulfill specific requirements.
   }
   {Our goal was to manufacture the mid-infrared integrated optics beam combiner for NOTT based on the double-Bracewell architecture and run a detailed laboratory characterization in the L$^{\prime}$ band. We focus in particular on the achievable raw and self-calibrated nulling ratios.
   }
   {We use a setup based on a double Michelson interferometer to produce four broadband coherent beams simulating the four telescopes of the VLTI and perform broadband nulling at room temperature. We also analyze the modal, chromatic, and polarization behavior of the integrated optics beam combiner, and measure its total throughput.
   }
   {We were able to manufacture a single-mode four-telescope double-Bracewell beam combiner in Gallium Lanthanum Sulfide mid-infrared transparent Chalcogenide glass using Ultrafast Laser Inscription. We show that the directional couplers forming the four-telescope beam combiner (4T-nuller) have an achromatic splitting ratio across the band 3.65 -- 3.85\,$\mu$m with a 40/60 and 50/50 splitting for the side couplers and the central coupler, respectively. We report a total throughput of 37\%, including the Fresnel losses that will be mitigated with anti-reflection coatings, and quantify differential birefringence. Operating at room temperature, with 200\,nm bandwidth centered at 3.8\,$\mu$m and without polarization control, we measure an average raw null of 8.13$\pm$0.03$\times$10$^{-3}$  and a self-calibrated null of 1.14$\pm$0.01$\times$10$^{-3}$. Finally, we show that a $\theta^6$ broad null can be experimentally reproduced in these conditions. This is, to our knowledge, the first measurement of a broadband L$^\prime$ deep null obtained with a four-telescope integrated optics beam combiner. 
   }
   {Following these promising results, the next step foresees testing the performance of the 4T-nuller in cryogenic conditions.
   }


   \maketitle

\section{Introduction} 
\label{sec:intro}
\noindent Indirect detection techniques \citep{Fischer2014} that have successfully revealed the bulk of the currently known exoplanet population have permitted to establish fundamental demographics properties \citep{Lissauer2023}. In the near future, an important objective in this field is to measure the atmospheric composition of a large sample of exoplanets, ultimately searching for bio-signatures \citep{DesMarais2002,Schwieterman2018}. As planets may form in the protoplanetary disk around young pre-main sequence stars, it is proposed that a high occurrence of Jupiter and sub-Jupiter planets should be found close to the water-ice line \citep{Fernandes2019,Fulton2021}, which marks a transition between volatiles-rich and volatiles-poor regions and consequently may influence the formation of large planets by core accretion. Probing the properties and composition of exoplanets in the water-ice line region via spectroscopy can then help us to constrain the planet formation mechanisms.\\
For performing spectroscopy of non-transiting exoplanets, direct detection techniques are indispensable \citep{Chauvin2023}. Such techniques face the limitations of a strong contrast between the planet and the parent star (typically between 10$^{-3}$ to 10$^{-6}$ depending on the planet mass, age and spectral band) and a small angular separation (in the range of 1 to 10\,mas at the expected separation of the water-ice line), resulting in the planetary signal being out-shined by the bright central star. 
The combination of high-angular resolution and high-contrast techniques, along with improved signal post-processing techniques, is key to overcome such limitations.\\
The two main techniques that address these challenges in the context of direct spectroscopy of exoplanets are coronagraphy \citep{Galicher2023} and nulling interferometry \citep{Bracewell1978, Burke1986, Angel1986,Angel1997}. Nulling is the prime technique for performing direct spectroscopy of close-in exoplanets in the mid-infrared L$^\prime$ band (3.5 -- 4.1 $\mu$m) since interferometric baselines can probe angular separations of 1-10\,mas not accessible with the ELT and VLT. At the same time, the rejection ratio achieved with nulling interferometry gives access to planet/star contrasts as low as $\sim$10$^{-4}$ \citep{Martinod2021, defrere+BTI+2016, hanot_pnf_nuller_2011}.\\
Detecting and characterizing exoplanets close to the water-ice line region is one objective of the NOTT\footnote{NOTT: \texttt{Nulling Observations of dust and planeTs}} visitor instrument at the VLTI \citep{Laugier2023,io_interferometry_asgard_hi5_2022} as part of the Asgard suite \citep{Martinod2023JATIS}. NOTT will perform spectroscopy of close-in giant exoplanets in the L$^{\prime}$ of L-band, a spectral sweet spot to detect the photons of self-luminous or irradiated close-in, young giant planets \citep{io_interferometry_asgard_hi5_2022}. As the first nulling instrument in the southern hemisphere, NOTT foresees to recombine the four beams of the VLTI in a self-calibrated nulling interferometric mode \citep{Martinache2018} to reach a contrast of $\sim$10$^{-4}$ at milliarcsecond separations. On the instrument side, NOTT will rely on a novel mid-infrared four-telescope (4T) integrated optics (IO) beam combiner. \\
IO beam combiners for interferometry favor an approach based on instrument miniaturization with improved opto-mechanical stability that has already been successfully developed and implemented at near-infrared wavelengths for both V$^{\rm 2}$ and nulling interferometry as in the case of GRAVITY in the K band \citep{io_interferometry_gravity_2018} and GLINT in the H band \citep{Martinod2021}, and now emerging at mid-infrared wavelengths \citep{Tepper2017,Gretzinger2019}.\\
In this paper, we report on the development of the first L$^{\prime}$ band, four-telescope, IO beam combiner that will perform nulling interferometry at the VLTI, also referred to as 4T-nuller. We recall the requirements of the integrated optics beam combiner, detail the design and manufacturing process, then report on the performance of the 4T-nuller and the level of extinctions measured in the laboratory. 

\section{Integrated optics beam combiner for the double-Bracewell scheme} \label{sec:dB_principle}
\begin{figure}[t]
    \centering
    \includegraphics[width=0.85\columnwidth]{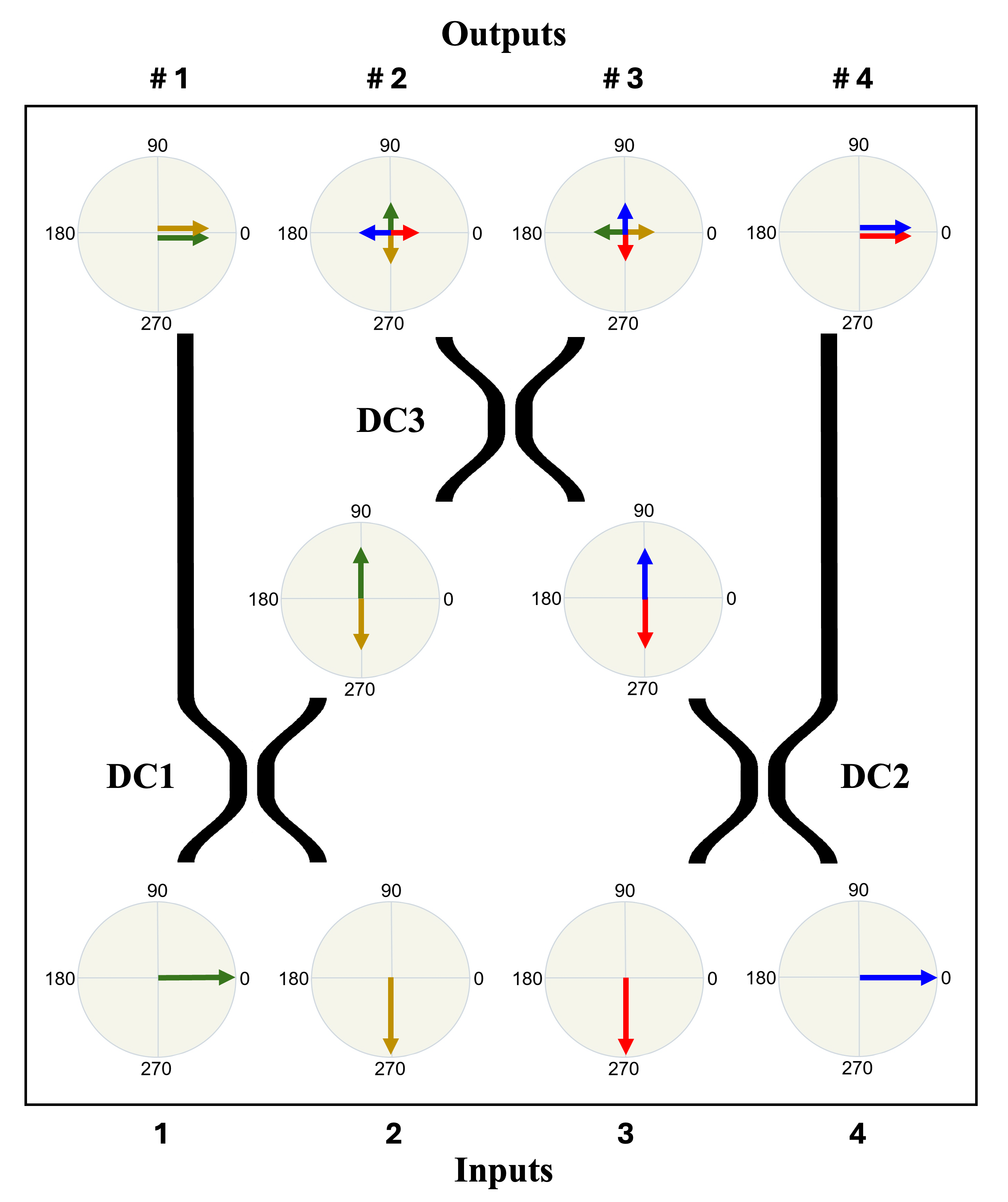}
    \caption{Functional phasor diagram of the double-Bracewell nuller with complex amplitudes of inputs 1 (green), 2 (yellow), 3 (red), and 4 (blue). 
	DC1, DC2, and DC3 are three directional couplers with 50/50 splitting ratio in intensity. 
    DC1 and DC2 produce constructive outputs at \#1 and \#4, while the two nulled outputs are feeding DC3. The nulled outputs \#2 and \#3 feature two transmission maps, which are centro-symmetric to one another w.r.t. the central star.
    }
	\label{fig:phasor}
\end{figure}

It is known that a typical two-telescope or single-Bracewell scheme \citep{Bracewell1978} with a $\theta^2$ null dependency - where $\theta$ is the on-sky angular direction - leads to stellar leakage coming from the partially resolved stellar that hampers the detection of the faint planetary signal. It is in principle possible to mitigate this effect by broadening the central null to $\theta^4$ or $\theta^6$ dependency, which requires interferometrically recombining four (or more) telescopes, as investigated by several authors \citep{Leger1996, Angel1997, Mennesson2005}. \\
The additional advantage of a multi-telescope beam combination resides in the implementation of internal modulation schemes. An example is the double-Bracewell architecture, which foresees two single-Bracewell nullers producing each two inherently centrally symmetric transmission maps w.r.t the on-axis central star, and which are then coherently combined with an additional relative phase shift of $\pi$/2 and -$\pi$/2, respectively. The results are two new transmission maps that lack individual centro-symmetry w.r.t the central star, but that are related to each other through centro-symmetry (e.g., Fig.~5 in \cite{Mennesson2005}). The modulation between the two nulled outputs allows to discriminate the planetary signal transmitted by the nuller from the transmitted signal of a centrally symmetric source, for instance, a circumstellar exo-zodiacal dust disk. \\
In the case of ground-based observations, the dominant source of errors in the null measurement is set by fast-varying optical path differences (OPD) due to fringe-tracking residuals. Interestingly, the double-Bracewell scheme previously described enables the implementation of a self-calibration technique in which the two nulled outputs are subtracted to produce a so-called kernel null that is less sensitive to residual phase excursions \citep{Martinache2018}. This architecture has been adopted for NOTT and assessed in terms of theoretical performance in \cite{Laugier2023}.\\
The IO version of the two-telescope single-Bracewell beam combiner can be implemented using a 3\,dB 2$\times$2 directional coupler with the transfer matrix $M$  given in Eq.~\ref{eq:dc_matrix}. 
\begin{eqnarray}
M =\begin{bmatrix}
\frac{\sqrt{2}}{2} & i\frac{\sqrt{2}}{2} \\
i\frac{\sqrt{2}}{2} & \frac{\sqrt{2}}{2}
\end{bmatrix}
\label{eq:dc_matrix}
\end{eqnarray}
The interaction length in the coupling region determines the level of splitting ratio (Fig. 1 in \cite{Tepper2017b}). For a 50/50 directional coupler, the constructive (anti-null) and destructive (null) outputs are obtained when an external $\pi$/2 phase shift is applied to one of the two in-phase incoming beams. 
Following \cite{Errmann2015}, the IO-based four-telescope double-Bracewell scheme can be obtained by cascading three 2$\times$2 couplers, as illustrated in Fig.~\ref{fig:phasor}.
\\
In the nulling configuration, the relative phases are 0, -$\pi$/2, -$\pi$/2, and 0 for inputs 1, 2, 3, and 4, respectively. The first stage contains two couplers that each combine the signals from a specific pair of telescopes. 
The input signals 1 and 2 are coupled to the first directional coupler (DC1), while the input signals 3 and 4 are directed to the second directional coupler (DC2). 
DC1 and DC2 split the signal intensities $I$ in half and introduce a relative phase shift of $+\pi/2$ on the cross-coupled beams, a characteristic inherent to waveguide directional couplers \citep{Foresto1995}. 
In the second stage, the resulting two nulled signals are recombined using a third coupler (DC3) that delivers two nulled outputs, \#2 and \#3. 
As a consequence of the input phase relationships, these two pairs also generate two nulled signals (stage 1) that can then be combined using the third central directional coupler DC3. Because DC3 also introduces a $\pi/2$ phase shift to the cross-coupled beam, we obtain at the nulled outputs \#2 and \#3 two transmission maps that are centro-symmetric to one another, hence reproducing the conditions for internal modulation in the "dual chopped Bracewell" scheme \citep{Mennesson2005}.
Simultaneously, the outputs \#1 and \#4 are two constructive states -- or anti-nulls -- obtained from each coupler DC1 and DC2, individually. \\
The constructive configuration complementary to the nulling one has relative phases of 0, +$\pi/2$, 0, and -$\pi/2$ for inputs 1, 2, 3, and 4, respectively. In this case, outputs \#1, \#3, and \#4 correspond to a destructive state, while output \#2 corresponds to a constructive state. Importantly, the sum of the fluxes from outputs \#1 and \#4 in the nulling configuration is equal to the flux from the output \#2 in the constructive configuration, assuming that all directional couplers have a 50/50 splitting ratio.\\
To estimate the achievable null depth, two nulling or extinction ratio quantities can be measured with the double-Bracewell IO beam combiner. The raw null $(rn)$ corresponds to the ratio of the destructive $I_-$ to the constructive $I_+$ output intensities, with the average raw null from \#2 and \#3 given by 
\begin{eqnarray}
rn = \frac{1}{2}\cdot\frac{I_2 + I_3}{I_1 + I_4} 
\label{eq:rawnull}
\end{eqnarray} 
\noindent The second quantity is the self-calibrated null ($scn$) defined by the difference between the two nulled outputs through 
\begin{eqnarray}
scn =\frac{|I_2 - I_3|}{I_1 + I_4} 
\label{eq:scnull}
\end{eqnarray}
\noindent The self-calibrated null is robust against upstream instrumental errors, such as differential piston and phase errors, as it represents the difference of complex conjugate electric fields, effectively canceling residual errors \citep{laugier2020, Cvetojevic2022, Chingaipe2023}.
As the depth of the self-calibrated null depends on the difference between the nulled outputs \#2 and \#3, it is critical to experimentally assess the symmetric behavior of the IO double-Bracewell.
	\begin{table}[t]
	\centering
	\renewcommand{\arraystretch}{1} 
	    \centering
	    \caption{Key requirements of the 4T-nuller.}
	    \scriptsize  
	    \begin{tabular}{lc}
	    \hline
		\hline
		\vspace{0.1cm}
		Property & Requirement \\ \hline		
			 Operational wavelength range & 3.5-4.0 $\mu$m \\
		   Modal behavior & Single-mode \\ 
			 Photometric splitting & 20/80\\ 
			 Interferometric splitting & 50/50 \\ 
   	     Splitting spectral response & Achromatic \\ 
	       Optical throughput (w/o Fresnel loss) & $\approx$50\% \\ 
    	   Null level (raw/self-cal.\tablefootmark{$^a$}) & 10$^{-2}$\,/\,10$^{-5}$\\ \hline
            \hline
            \end{tabular}
            \tablefoot{
            \tablefoottext{$^a$}{The self-calibrated null $\sim10^{-5}$} corresponds to the state after post-processing, as described by \cite{hanot_pnf_nuller_2011} and \cite{Mennesson2011}, and is not considered in this work.}
        \label{tab:photonic_requirements}
\end{table}
\section{Development of NOTT's integrated optics beam combiner}\label{sec:nott_requirments}
\subsection{Requirements and design}
\label{sec:nott_req} 	
The high-level requirements of the IO beam combiner, derived primarily from the scientific requirements of NOTT \citep{Defrere2024}, are reported in Table~\ref{tab:photonic_requirements}. It shall be operational for the L$^\prime$ band, and the waveguides shall exhibit single-mode behavior over the considered bandwidths for the purpose of wavefront filtering \citep{Ruilier2001}. To minimize the degradation of the broadband raw null due to photometric imbalance, an achromatic splitting ratio across the L$^\prime$ band is required. A minimum throughput of $\sim$50\% shall be reached after reduction of the Fresnel losses.
\begin{figure}[t]
    \centering
    \includegraphics[width=0.95\columnwidth, angle=0]{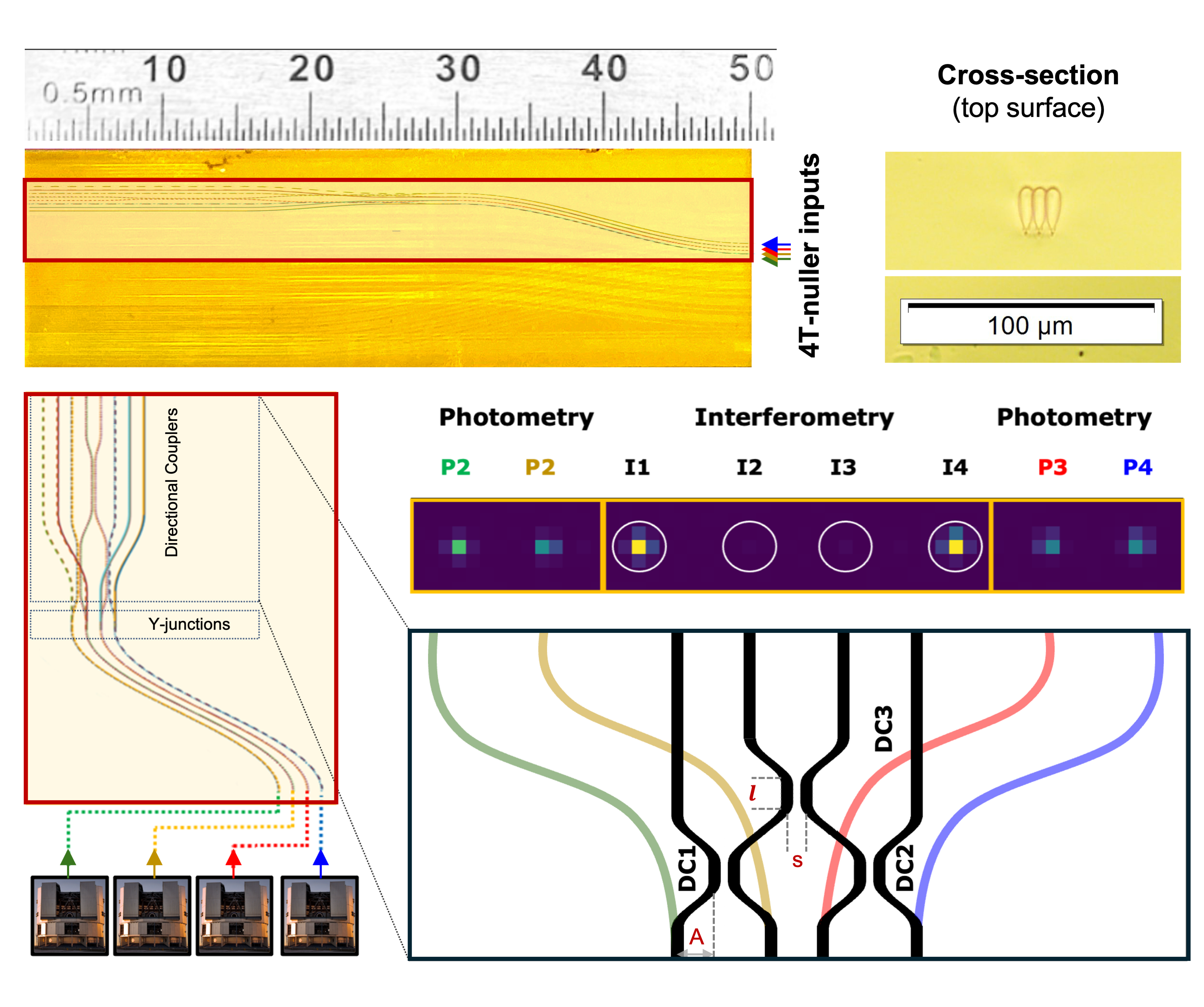}
    \caption{
    Top left: NOTT's 4T-nuller photonic chip with the network of waveguides manufactured by ULI (see Sect.~\ref{sec:manufacturing_uli} for details). Top right: cross-section of the input/output facet of a triplet waveguide; Bottom left: schematic of the 4T-nuller showing the four inputs, the side-step S-bends, and the three cascaded directional couplers; Bottom right: visualization of the different outputs of the 4T-nuller (photometry and interferometry) observed on the testbench camera (Sect.~\ref{sec:results_extinction_ratios}). The photometric signals from each input are extracted by the Y-junctions and routed to photometric outputs $P_1$, $P_2$, $P_3$, and $P_4$. The remaining signals pass through the interferometric section consisting of three directional couplers, producing the four circled interferometric outputs $I_1$, $I_2$, $I_3$, and $I_4$. In the configuration of Fig.~\ref{fig:phasor}, $I_1$ and $I_4$ correspond to constructive outputs while $I_2$ and $I_3$ yield isolated nulled signals. {$l$ is the interaction length of the coupler, $s$ is the center-to-center separation between the waveguides or pitch, $A$ is the amplitude of the lateral offset and $r$ is the radius of curvature.}
    }
    \label{fig:4T-nuller-layout}
\end{figure}\\
\begin{figure*}[t]
    \centering
    \includegraphics[width=0.8\textwidth]{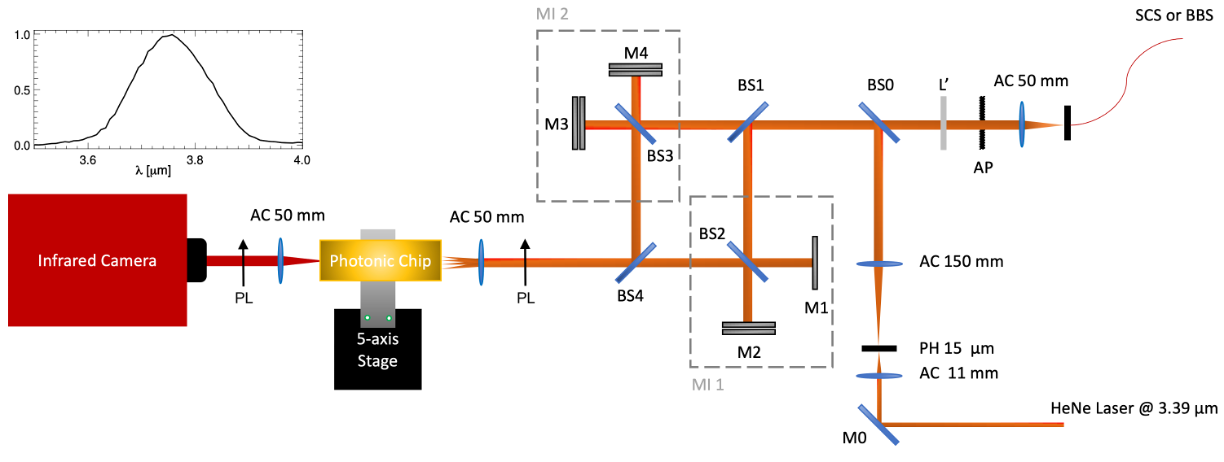}
    \caption{Layout of the mid-infrared characterization testbench and its components. SCS: supercontinuum source; BBS: blackbody source; AC: achromat; AP: aperture; L$^\prime$: broadband filter; PH: pinhole; BS: beam splitter; M1: fixed mirror; M2, M3, M4: movable mirrors with delay lines; PL: mid-infrared linear polarizer; HeNe: metrology laser to calibrate the delay lines; Photonic Chip contains the 4T-nullers. Top inset: spectral profile of the SCS measured before the photonic chip.}
    \label{fig:bench}
\end{figure*}
The design of the NOTT's IO beam combiner, 4T-nuller, follows the layout shown in Fig.~\ref{fig:4T-nuller-layout}. 
The four VLTI telescopes are coupled to the four input waveguides of the 4T-nuller spaced by 125\,$\mu$m using the fore-optics of NOTT \citep{Garreau2024}. 
Side-step S-bend waveguides are implemented after the inputs, before the beam combination region, to minimize the contamination of the output waveguides by uncoupled stray light that may propagate in the substrate \citep{Norris2014}. 
Y-junctions are implemented for the beam splitting function, routing 80\% of the signal towards the directional couplers while 20\% of the signal is routed to the photometric taps for flux monitoring and photometric calibration. 
The beam combination function is ensured by three cascaded 50/50 directional couplers following the Angel\,\&\,Woolf scheme, as described in Sect.~\ref{sec:dB_principle}. 
This design delivers four photometric outputs ($P_{\#}$) and four interferometric outputs ($I_{\#}$). The spacing between the outputs is optimized for the NOTT’s back-end optical design with: 
190\,$\mu$m between $P_1$ and $P_2$; 
200\,$\mu$m between $P_2$ and $I_1$; 
240\,$\mu$m between $I_1$ and $I_2$; 
200\,$\mu$m between $I_2$ and $I_3$; 
240\,$\mu$m between $I_3$ and $I_4$; 
200\,$\mu$m between $I_4$ and $P_3$; 
190\,$\mu$m between $P_3$ and $P_4$. 
This non-equidistant spacing prevents overlap between the various output signals, ensuring that the nulled outputs are sufficiently isolated to avoid cross-talks between the constructive/anti-null and photometric signals on the NOTT's detector.

\subsection{Fabrication} 
\label{sec:manufacturing_uli}
Following the studies of \cite{arriola_delN_changes2017mid,Tepper2017,Gretzinger2019}, the 4T-nuller of NOTT was manufactured using the process of Ultrafast Laser Inscription (ULI) \citep{Nolte2003,Thomson2009,Gross2015} in Gallium Lanthanum Sulphide (GLS) chalcogenide glass with a refractive index n\,$\approx$\,2.31 at 3.4\,$\mu$m and a high transparency from 0.5 to 9\,$\mu$m.
We implemented the multiscan technique using a Ti:Sapphire femtosecond laser operating at 800\,nm and delivering pulses of $<$50\,fs duration at a repetition rate of 5.1\,MHz. These parameters allow working in the thermal fabrication regime where local changes in the refractive index of the glass occur by cumulative heating \citep{eaton2005}. Based on the triple-track waveguides (inset of Fig.~\ref{fig:4T-nuller-layout}) the Y-junctions and directional couplers of the 4T-nuller were formed. For the interferometric functionality, we implemented asymmetric directional couplers to obtain a close to achromatic splitting ratio across the L$^{\prime}$ bandwidth. This was done by altering the propagation constant, i.e. the effective index, of the guided mode in the right waveguide within the interaction region of the coupler. The change in effective index was achieved by reducing the writing velocity from 100\,mm/min to 55\,mm/min while maintaining a pitch $s$ of 26\,$\mu$m between the waveguides. \\
Using the ULI platform, we fabricated through several iterations (cf. Sect.~\ref{sec:dir_couplers}) the 4T-nuller specified in Sect.~\ref{sec:nott_req} in a slab or photonic chip with dimensions of $\sim$50\,mm in length, 15\,mm in width and a thickness of 2\,mm after grinding and polishing of the input/output facets (Fig.~\ref{fig:4T-nuller-layout}). The 4T-nuller inputs are located 180\,$\mu$m below the top surface of the photonic chip. For the side-step S-bend waveguides, the bending losses were minimized by adopting a minimum radius of curvature of 45\,mm and an amplitude of the lateral offset of 1.1\,mm. Additionally, the S-bends used in this 4T-nuller are cosine S-bends with varying radii of curvature due to their minimal bending loss properties \citep{kruse2015, Gretzinger2019}. In the interferometry section emphasized in Fig.~\ref{fig:4T-nuller-layout}, the couplers DC1 and DC2 have a cosine S-bend offset amplitude $A \approx 0.05$\,mm, a minimum radius of curvature $r$=50\,mm, and a total S-bend length of 3.39\,mm.  For DC3, the parameters are $A\approx$0.1\,mm, $r$=50\,mm, and the total length is 4.64\,mm. 
\subsection{Testbed}
\label{sec:setup}
To experimentally assess the performance of our 4T-nuller, we use the testbed described in \cite{Tepper2018} that generates four coherent beams and includes phase delay capabilities (Fig.~\ref{fig:bench}). The testbed utilizes two cascaded Michelson interferometers that create four beams, which can then be independently steered in front of the input waveguides by tilting any of the $M1$, $M2$, $M3$, and $M4$ mirrors. Using four pellicle beam splitters, $BS1$ to $BS4$ for the 3--5\,$\mu$m range minimizes the chromatic dispersion otherwise introduced by thick beam splitters. The high symmetry of the arrangement ensures the delivery of four beams with equal intensity. In practice, we observe small discrepancies of approximately 10--15\% in the flux level, likely due to small differences in the absolute transparency of each pellicle beam splitter. The input light sources can be chosen between a fiber 2--5\,$\mu$m mid-infrared super-continuum white light laser from  Leukos or a fiber blackbody source at $T$=1500\,K from Thorlabs. The sources are routed using a mid-infrared single-mode InF$_3$ fiber, whose output is collimated to feed the testbed. Three servomotors connected to the linear stages supporting the mirrors $M2$, $M3$, and $M4$ scan the optical path delay.
\begin{figure*}[t]
    \centering
    \includegraphics[width=1.0\textwidth]{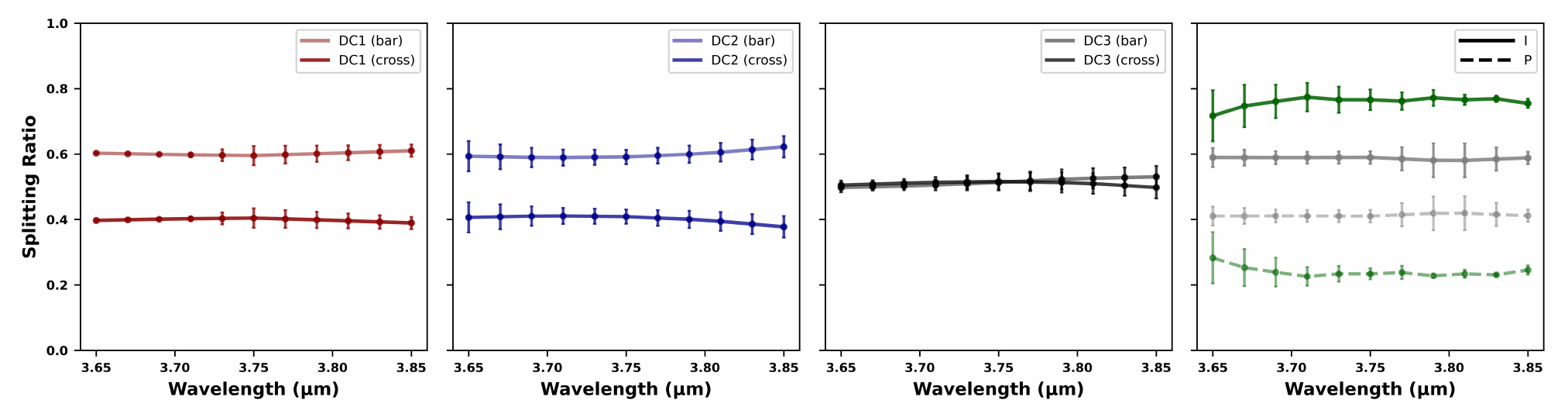}
    \caption{Measurement of the splitting ratio of the three directional couplers with 7.5\,mm interaction length forming the 4T-nuller (three plots from the left) and for the Y-junction photometric tap. The error bars correspond to three times the standard deviation obtained from four independent measurements. In the last plot, the gray curves depict one of the four Y-junctions in the characterized 4T-nuller, while the green curves show the properties of the targeted 80/20 splitter, which will be implemented in a future version.}
    \label{fig:4T-nuller-DCs-splitting}
\end{figure*}\\
Free-space mid-infrared achromatic doublets with focal length $f=50$ mm are used to couple light in and out of the 4T-nuller from the edge of the photonic chip. It is positioned on a manual 5-axis micrometric positioning stage with differential drives that includes the MBT402D/M stage, which allows us to precisely align the waveguides to the input beam. 
The signal is recorded using an InSb Infratech camera. The setup can be used in Fourier Transform Spectroscopy (FTS) mode to characterize the spectral behavior of the 4T-nuller. For this purpose, the HeNe laser at 3.39\,$\mu$m delivers the metrology to calibrate the optical path delay. The spectral profile of the source with the employed L$^\prime$ bandpass filter is presented in the inset of Fig.~\ref{fig:bench}.

\section{Results}
This section reports the results of the laboratory characterization assessing the performances of the 4T-nuller and the properties relevant for nulling interferometry in the L$^{\prime}$ band. 
\subsection{Modal behavior}
\label{sec:modal_behavior} 
Near-field imaging was conducted using a 3.39\,$\mu$m laser and a microscope objective with 18\,mm focal length with 12.5$\times$ magnification. The Mode-Field Diameters (MFDs) were extracted using Gaussian fitting. 
The measurements revealed that the average MFDs, taken from both horizontal and vertical cross-sections, ranged between 26--28\,$\mu$m, with a circularity of $0.95 \pm 0.03$ \citep{sanny_phd2024}. Furthermore, the mode-profile remained unchanged when displacing the injection spot laterally, confirming the single-mode property at 3.39\,$\mu$m. In the case of the L$^\prime$ band, variations in the injection conditions similarly affected all output intensities, indicating a single-mode behavior of the 4T-nuller.
\subsection{Splitting ratios} 
\label{sec:results_splitting_DCs}
\subsubsection{Directional couplers}
\label{sec:dir_couplers}
We first manufactured  34 2$\times$2 asymmetric directional couplers in GLS glass to identify the ULI parameters required for achromatic behavior of the splitting ratio. The parameters optimization focused on optimal interaction length $l$ and the triple-track writing order (left-to-right vs. right-to-left)\footnote{In the multiscan technique, the writing order of the triple-track or triplet forming a single waveguide refers to the order or sequence in which the left, central, and right track is inscribed \citep{Sanny_2022}.}. 
Different interaction lengths $l$ ranging from 6.0 to 8.0\,mm in steps of 0.5\,mm were tested. In this phase, we also assessed the reproducibility of the coupler's behavior in terms of splitting ratio. \\
Using the FTS mode of our testbed, we measured the spectral splitting ratios of the couplers across a 3.65 -- 3.85\,$\mu$m wavelength range. This was performed by injecting light in the left or right input waveguide of the coupler and measuring the bar (i.e., same side of the excited input) and cross (i.e., opposite side of the excited input) flux power $P_{\rm cross}$ and $P_{\rm bar}$, respectively. For each spectral channel, the normalized ratios $P_{\rm cross}$/($P_{\rm cross}$+$P_{\rm bar}$) and $P_{\rm bar}$/($P_{\rm cross}$+$P_{\rm bar}$) are retrieved. 
The parameter optimization showed that 2$\times$2 directional couplers with $l$=7.5\,mm interaction length and left-to-right track-writing order exhibited close to 50/50 achromatic splitting. In order to confirm this behavior, five 4T-nullers containing three cascaded couplers DC1, DC2, and DC3 were manufactured, with each also having an interaction length $l$ varying from 6.0 to 8.0 mm in steps of 0.5\,mm.\\
The spectral characterization showed that the 4T-nuller with $l$=7.5\,mm resulted in an achromatic and balanced 50/50 splitting ratio for the central coupler DC3 (Fig.~\ref{fig:4T-nuller-DCs-splitting}) in agreement with the results from the reference 2$\times$2 couplers. Conversely, DC1 and DC2 exhibited an achromatic but unbalanced splitting ratio (Fig.~\ref{fig:4T-nuller-DCs-splitting}). 
Interestingly, the 4T-nuller with $l$=6.5\,mm showed a balanced 50/50 achromatic splitting ratio for DC1 and DC2, but an unbalanced achromatic splitting ratio for DC3. 
We anticipate that this discrepancy could result from the long-range stresses  \citep{ULI_del_beta2018} during the laser writing process. For a future improvement, our findings suggest an interaction length $l_{\rm 1,2}$=6.5\,mm and $l_{\rm 3}$=7.5\,mm for DC1/DC2 and DC3, respectively.\\
Importantly, we find that, despite different values of splitting ratio, all couplers DC1, DC2 and DC3 show a spectrally flat splitting ratio. Since the central combiner DC3 is responsible for extracting the self-calibrated null, we selected the 4T-nuller with $l$=7.5\,mm that ensures a balanced splitting ratio to pursue the evaluation of the interferometric performances.
\subsubsection{Y-junctions} 
\label{sec:results_splitting_Y_junctions}
The four 1$\times$2 Y-junctions used in this work as photometry taps, as shown schematically in Fig.~\ref{fig:4T-nuller-layout}, split a waveguide into two arms via a tapered transition region, which gradually changes the dimensions of the waveguide cross-section \citep{Y_Junction_operation_single_mode1982}. To prevent any physical crossing between waveguides, we exploited the unique property of ULI that enables three-dimensional photonic architectures and routed the photometric taps about 60\,$\mu$m underneath the interferometric waveguides, before gradually rising them to the same horizontal plane as the interferometric outputs. 
Similar to the directional couplers, we characterized the spectral splitting ratio of the Y-junctions in the 4T-nuller and found an achromatic ratio of 40/60 for the photometric taps and the interferometric channel, respectively. Subsequently, we modified the geometry of the Y-junction \citep{sanny_phd2024} and experimentally assessed that the required 20/80 ratio can be achieved with less than 5\% deviation from a purely achromatic behavior between 3.65\,$\mu$m and 3.85\,$\mu$m (Fig.~\ref{fig:4T-nuller-DCs-splitting}). The newly engineered Y-junction will be included in the fabrication of a future 4T-nuller.
\subsection{Throughput} 
\label{sec:results_throughput}
To measure the total throughput of the 4T-nuller, we carefully optimized the injection of the input beams into one or several inputs. This was achieved by maximizing the output flux on the camera and quantifying the total flux emerging from all outputs. This is then compared to the total injected flux by removing the sample in front of the injection optics to derive a relative flux ratio. The high temporal stability of the super-continuum white light laser source over the timescale of the measurement is expected to have a negligible impact on the uncertainty of the reported throughput. The throughput of the 4T-nuller was also compared with six straight and five side-step S-bends reference waveguides written on the same photonic chip to investigate the impact of the bending losses. 
\begin{figure}[t]
    \centering
    \includegraphics[width=0.90\columnwidth]{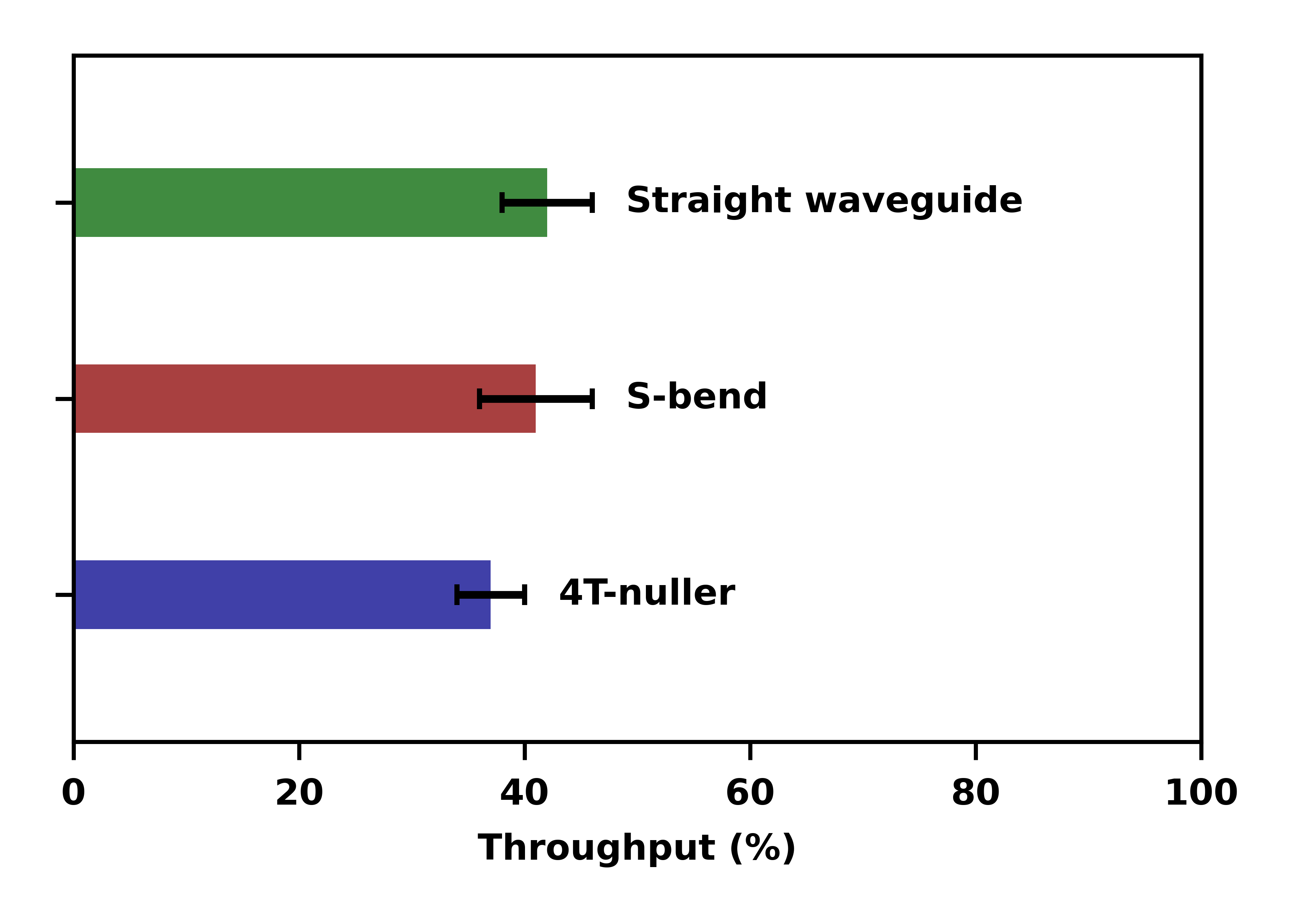}
    \caption{Relative throughput in percent for the reference straight waveguide and S-bends of 1.1 mm amplitude, along with the 4T-nuller. Error bars denote standard deviations from multiple identical straight and S-bend waveguides on the same photonic chip, while the standard deviation for the 4T-nuller is based on the throughputs of its four individual inputs.
    }
    \label{fig:throughput}
\end{figure}
The results of Fig.~\ref{fig:throughput} show that straight waveguides exhibit a total throughput of 42\%\,$\pm$\,4\%, and the side-step S-bends a total throughput of 41\%\,$\pm$\,5\%, suggesting a low impact bending losses. The uncertainties are estimated from the standard deviation in the measurement obtained for the different waveguides of each type. \\
The 4T-nuller, which incorporates side-step S-bends with the same radius of curvature and offset amplitude as the reference S-bends, features a throughput $T$\,=\,37\%\,$\pm$\,3\%. The small throughput degradation results likely from the additional (bending) losses due to the Y-junctions and the cascaded directional couplers. \\
\noindent The throughput $T$ can also be reported in terms of insertion losses $I_L$=10$\log_{10}$($T$) in dB, which includes the Fresnel reflection losses $F_L$ at the two facets of the chip, the coupling losses $C_L$ resulting from the mode-mismatch between waveguide and injection spot, and the internal losses $P_L$ inherent to the waveguide design and manufacturing imperfections including the bending losses, propagation losses and transition losses from the Y-junction splitters. The coupling loss $C_L$, based on the $\sim$17\,$\mu$m 1/e$^2$ diameter of the injection spot and the $\sim$27\,$\mu$m mode-field-diameter (MFD) of the fundamental mode, is estimated to be 18.6\% or -0.9\,dB. The GLS glass refractive index of $n$=2.36 at 4\,$\mu$m induces a 15.4\% reflection per facet or -0.73\,dB loss. Using the relation $P_L$ = $(1/L_{chip})(I_L-2 F_L-C_L)$ and a total average chip length of $L_{chip} = 50$ mm, the propagation losses are found to be $P_L$=-0.30$\pm$0.09\,dB/cm for the S-bends and 
$P_L$=-0.28$\pm$0.08\,dB/cm for the straight waveguide, respectively.\\ 
This result can be compared to the propagation losses for GLS laser-written straight waveguides of $-0.22\pm0.02$\,dB/cm at 4\,$\mu$m reported by \cite{Gretzinger2019}. Such comparable propagation loss furthermore indicates the reproducibility of similar waveguides. The marginal difference between the reference straight and S-bend waveguides also indicates that the presence of well-designed S-bend waveguides and photonic transitions in the 4T-nuller does not dramatically degrade the overall performance. \\
As part of a future improvement, it is anticipated that the deposition of anti-reflection coatings onto the photonic chip's facets will further improve the total throughput of the device. Limiting the reflection losses to a conservative value of 2\% per facet would increase the total throughput to $\sim$49\%, close to the original requirement (Table \ref{tab:photonic_requirements}).
\subsection{Polarization properties}
\label{sec:results_polar_properties}
\begin{figure}[t]
    \centering
\includegraphics[width=1.0\columnwidth]{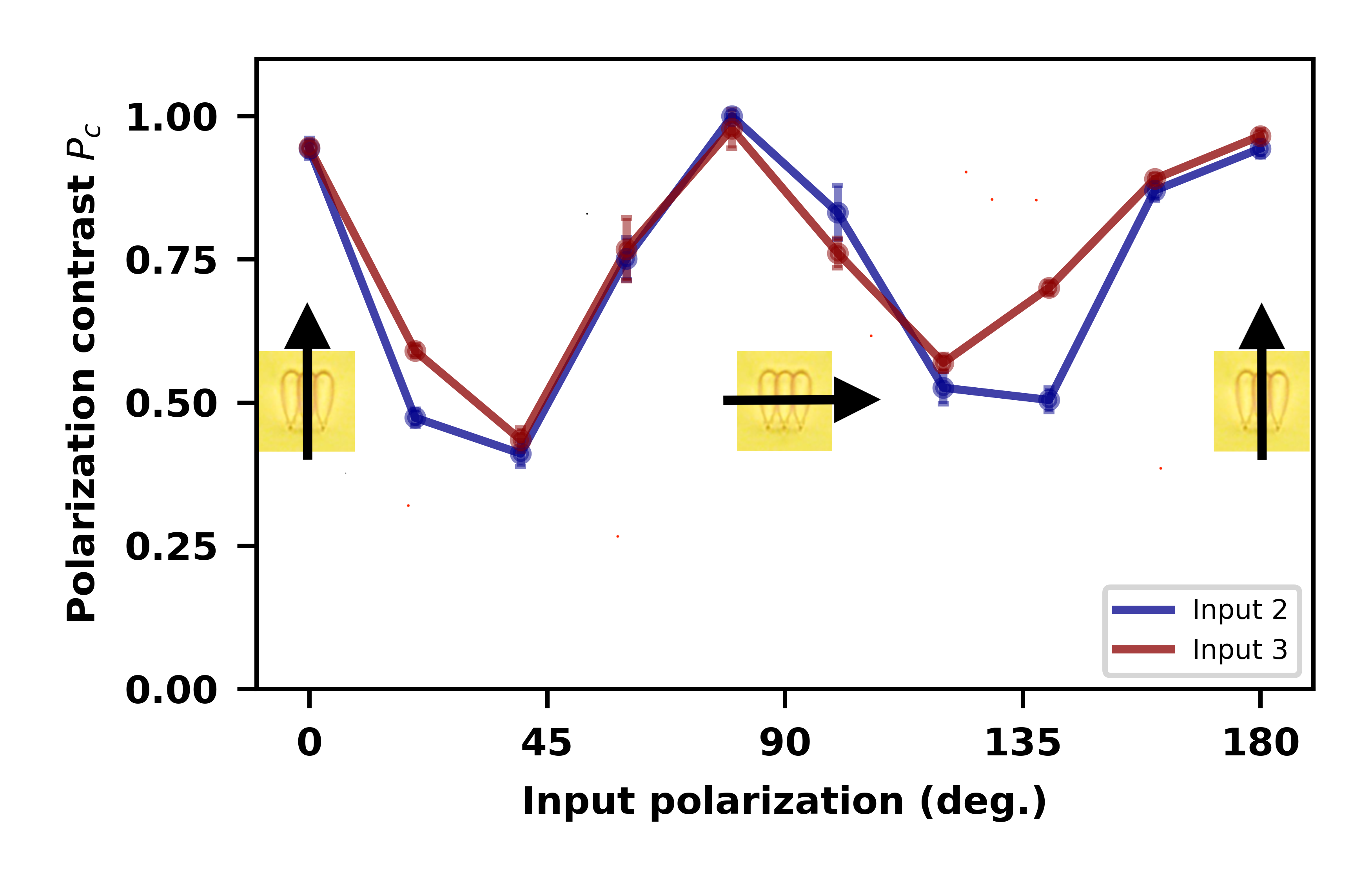}
    \caption{Polarization contrast at the 4T-nuller output $I_2$ as a function of the input polarization angle at inputs 2 and 3 in steps of 20$^{\circ}$. The error bars show the standard deviation over 200 points for each input polarization orientation. The arrows indicate the orientation of the polarized beam relative to the vertical position of a triple-track waveguide’s cross-section in the inset.}
    \label{fig:pola_contrast}
\end{figure}

\begin{figure}[t]
    \centering
\includegraphics[width=1.0\columnwidth]{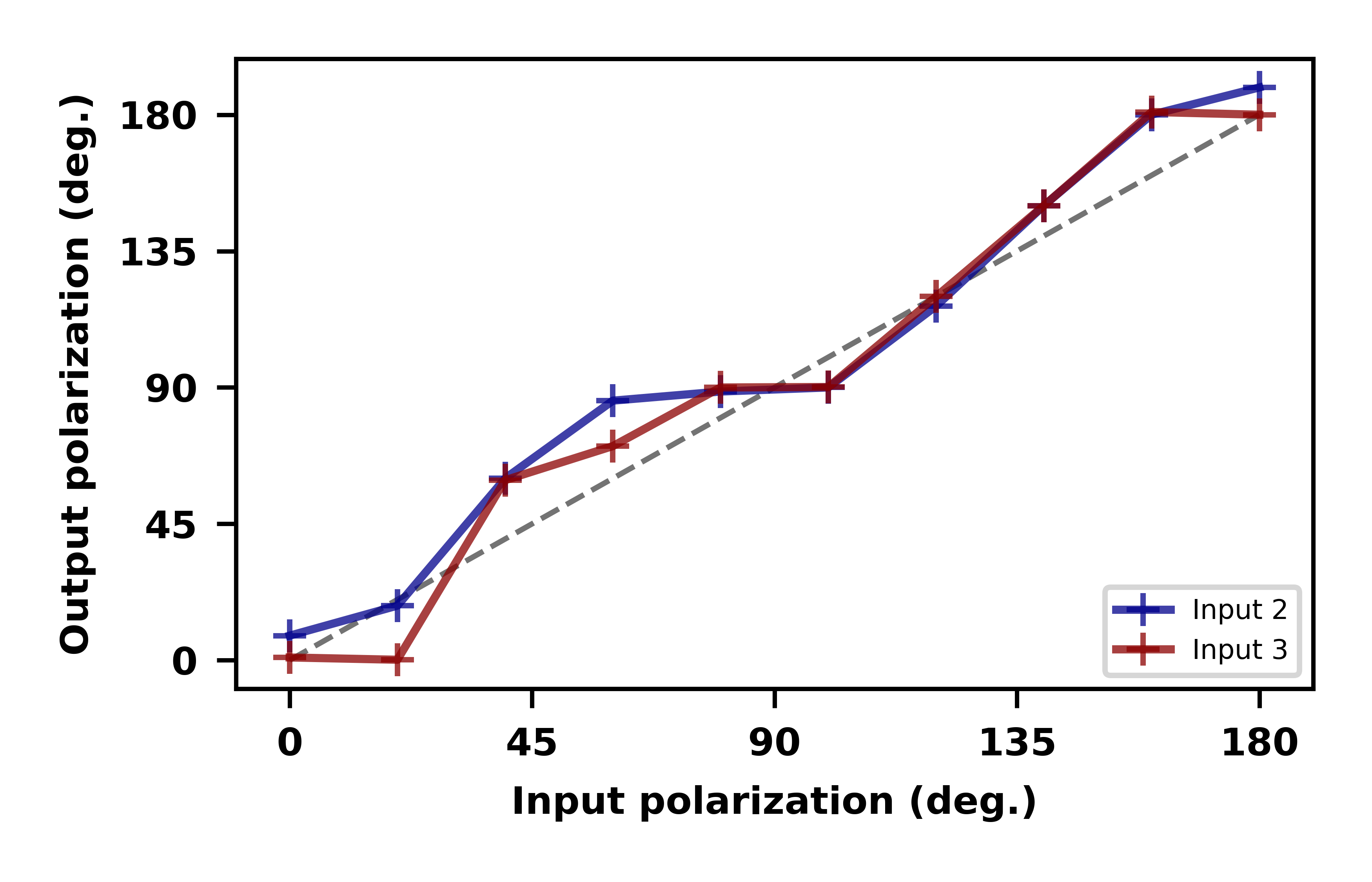} 
    \caption{Orientation of the output elliptical polarization measured at $I_2$ as a function of the angle of the linear input polarization at inputs 2 (blue) and 3 (red). The dashed line corresponds to the case where the input polarization angle is maintained at the output.}
    \label{fig:pola_angle}
\end{figure}

The interferometric contrast can be degraded by polarization mismatches between the beams as a result of differential stress or birefringence. Thus, it is important to assess and quantify the effect of differential birefringence as presented in this section.\\
We analyzed how the potential waveguide birefringence may induce a change in the polarization state of a given input linear polarization. We investigated at (1) whether the input linear polarization is maintained or modified to an elliptical polarization (cf. Fig.~\ref{fig:pola_contrast}); (2) the angular orientation of the output polarization state w.r.t. the input polarization (cf. Fig.~\ref{fig:pola_angle}). \\
\noindent For this purpose, a mid-infrared polarizer with linearly spaced wire grids (Thorlabs WP25M-UB) is integrated upstream of the 4T-nuller injection lens to set the angle of the input linear polarization. A second wire grid polarizer, used as an analyzer, is located downstream of the output collimation lens. The polarization state is probed at the outputs $I_2$ and $I_3$, with input linear polarization angles ranging between $0^{\circ}$ and $180^{\circ}$. \\
Fig.~\ref{fig:pola_contrast} shows the polarization contrast $P_c$ measured at the output $I_2$ as a function of the angle of the input linear polarization at inputs 1 and 2.  The contrast is estimated with Eq.~\ref{eq:contrast}, where $I_{min}$ and $I_{max}$ are the measured minimum and maximum intensity values for a given angle of the polarizer. 
\begin{eqnarray}
P_c = \frac{I_{max}-I_{min}}{I_{max}+I_{min}}
\label{eq:contrast}
\end{eqnarray}
In the two extreme cases, $P_c$=1 and $P_c$=0 correspond to a linear and circular polarization, respectively. The graph clearly shows the effect of the waveguide birefringence for specific orientation of the input polarization, with a contrast as low as 0.4 at $\sim$40$^\circ$ and 130$^\circ$ indicative of an elliptical polarization, whereas the contrast is close to unity for 0$^{\circ}$, 90$^{\circ}$, and 180$^{\circ}$.
\\
Fig.~\ref{fig:pola_angle} shows that the linear polarization angle set at the inputs 1 and 2 is well maintained at output $I_2$ and almost unchanged close to 0$^{\circ}$, 90$^{\circ}$, and 180$^{\circ}$. {The maximum mismatch is $\sim$18$^{\circ}$ for 20$^{\circ}$ polarization angle, while the average mismatch is $6^{\circ} \pm 2 ^{\circ}$.} A very similar result is found when measuring these effects at output $I_3$. \\
Importantly for interferometric measurements, the behavior is found to be very similar in trend between inputs 2 and 3, which points to low differential birefringence between the waveguides. This behavior also occurs in other IO beam combiners manufactured using ULI in different materials \citep{uli_birefringence2012,Benoit2021,Siliprandi2024}, suggesting possible shape birefringence induced by the laser writing process.

\subsection{High-contrast nulling performance}
\subsubsection{Nulling ratios} 
\label{sec:results_extinction_ratios}
A central assessment of the performance of the IO beam combiner in the context of nulling is the experimental estimate of the nulling ratio. For this purpose, we aim to reproduce in the lab the beam combination scheme identified in the phasor diagram of Fig.~\ref{fig:phasor}. The goals are twofold: measuring the nulls at the outputs \#2 and \#3, and measuring the self-calibrated null obtained from the difference between the outputs \#2 and \#3. \\
The experiment is conducted under the following conditions: (1) we operate in an air-conditioned laboratory at room temperature ($\pm$\,1$^{\circ}$C stability); (2) we operate in broadband conditions using the super-continuum source with the spectral bandpass shown in Fig.~\ref{fig:bench} and without polarization control; (3) the IO beam combiner or 4T-nuller tested for nulling ratio does not incorporate the final 20/80 photometric taps yet. However this is uncritical for this task since the classical calibration of photometric imbalance is not required here; (4) the beam injected into input 1 serves as a reference, while the other three beams for inputs 2, 3, and 4 are phased-up each by independent delay lines $M2$, $M3$, $M4$; (5) the delay lines have a minimum resolution of $\sim$29\,nm, limiting our ability to fine tune the delay.
\begin{figure}[t]
    \centering
    \includegraphics[width=0.95\columnwidth]{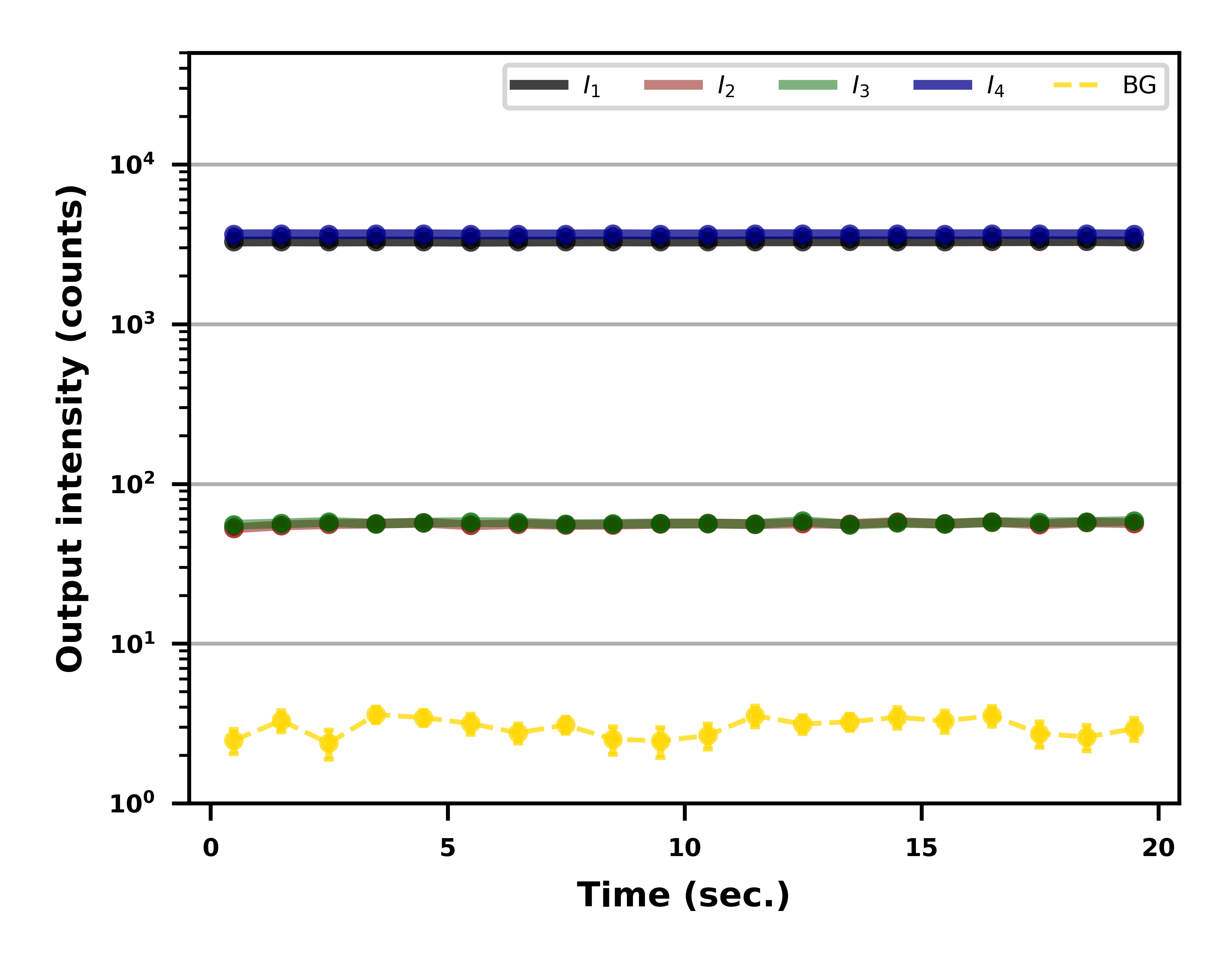}
    \caption{Flux, in counts (camera), for the constructive outputs $I_1$ and $I_4$ and the null depths as destructive outputs $I_2$ and $I_3$. The yellow line shows the residual contribution of the thermal background estimated from the image background in Fig.~\ref{fig:4T-nuller-layout}.
    }
    \label{fig:self_calibrated_outputs}
\end{figure}
\\
The measurement procedure is implemented as follows. 
The first step consists in setting all the telescope beams to the zero OPD position, which is identified as the center of the broadband sine-wave fringe packet. Following Fig.~\ref{fig:bench}, we first adjust each input individually to achieve the same photometric level for the output pair $I_2$ and $I_3$ by appropriately and slightly decoupling the corresponding input beam, as the testbed does not provide four strictly equal intensity beams. Secondly, we co-phase inputs 1 and 2, then 2 and 3, and finally 3 and 4. This procedure ensures that 50/50 DC3 (Fig.~\ref{fig:4T-nuller-DCs-splitting}) is delivered in a co-phased manner and that all inputs yield a comparable level of intensity, which constitutes a critical requirement for the self-calibrated null, as illustrated in Fig.~\ref{fig:phasor}.\\
Next, the inputs 3 and 4 in Fig.~\ref{fig:phasor} are blocked. M2 is delayed by -$\pi$/2, which is obtained by maximizing the flux at the output $I_1$ (constructive state) and minimizing the flux at outputs $I_2$ and $I_3$ (destructive state). Then, beams 3 and 4 are unblocked while 1 and 2 are blocked. The same procedure is applied where M3 is delayed by -$\pi$/2 to maximize the flux at output $I_4$ and minimize the flux at the outputs $I_2$ and $I_3$. Finally, beams 1 and 2 are unblocked. With the four beams injected, we then experimentally obtain the different interferometric states measured in Fig.~\ref{fig:4T-nuller-layout}. Because the optical magnification of the output imaging lens is unity, the measured fluxes are mostly concentrated to single camera pixels with 30\,$\mu$m size. The measured flux levels are reported in Fig.~\ref{fig:self_calibrated_outputs} and we can observe that the constructive outputs $I_1$ and $I_4$ and the destructive outputs $I_2$ and $I_3$, respectively, show almost identical levels of flux.\\
Finally, we compute the raw null and the self-calibrated null as described in Sect.~\ref{sec:dB_principle}. We emphasize that, because of DC1 and DC2 having an imbalanced splitting ratio of 40/60 instead of 50/50, the sum $I_1$+$I_4$ in the nulling configuration is not strictly equal to $I_2$+$I_3$ in the constructive configuration. Using the known splitting ratios of DC1 and DC2, we calculate that the average raw (Eq.~\ref{eq:rawnull}) and self-calibrated (Eq.~\ref{eq:scnull}) nulls derived from the data of Fig.~\ref{fig:self_calibrated_outputs} shall be corrected according to $rn$=$\frac{1}{2}$$\frac{(I_2 + I_3)}{0.96(I_1 + I_4)}$ and $scn = \frac{|I_2 - I_3|}{0.96(I_1 + I_4)}$ respectively, as described in Appendix~\ref{annex1}. The results presented in Fig.~\ref{fig:exp_rawnull_scnull} show a time averaged raw null of \textcolor{black}{8.13$\pm$0.03$\times$10$^{-3}$} over 20 seconds, while the time averaged self-calibrated null is \textcolor{black}{1.14$\pm$0.01$\times$10$^{-3}$} over 20 seconds. \\
We compare these nulling ratios to the dynamic range of our experiment, which is estimated to be 4.8$\times$10$^{-4}$ of the constructive state 0.96($I_1$+$I_4$). 
The comparison shows that the raw null is less than the dynamic range of our experiment, whereas the self-calibrated null is much closer to it. Possible causes are discussed in Sect.~\ref{sec:discussion}.
\begin{figure}[t]
    \centering
    \includegraphics[width=0.95\columnwidth]{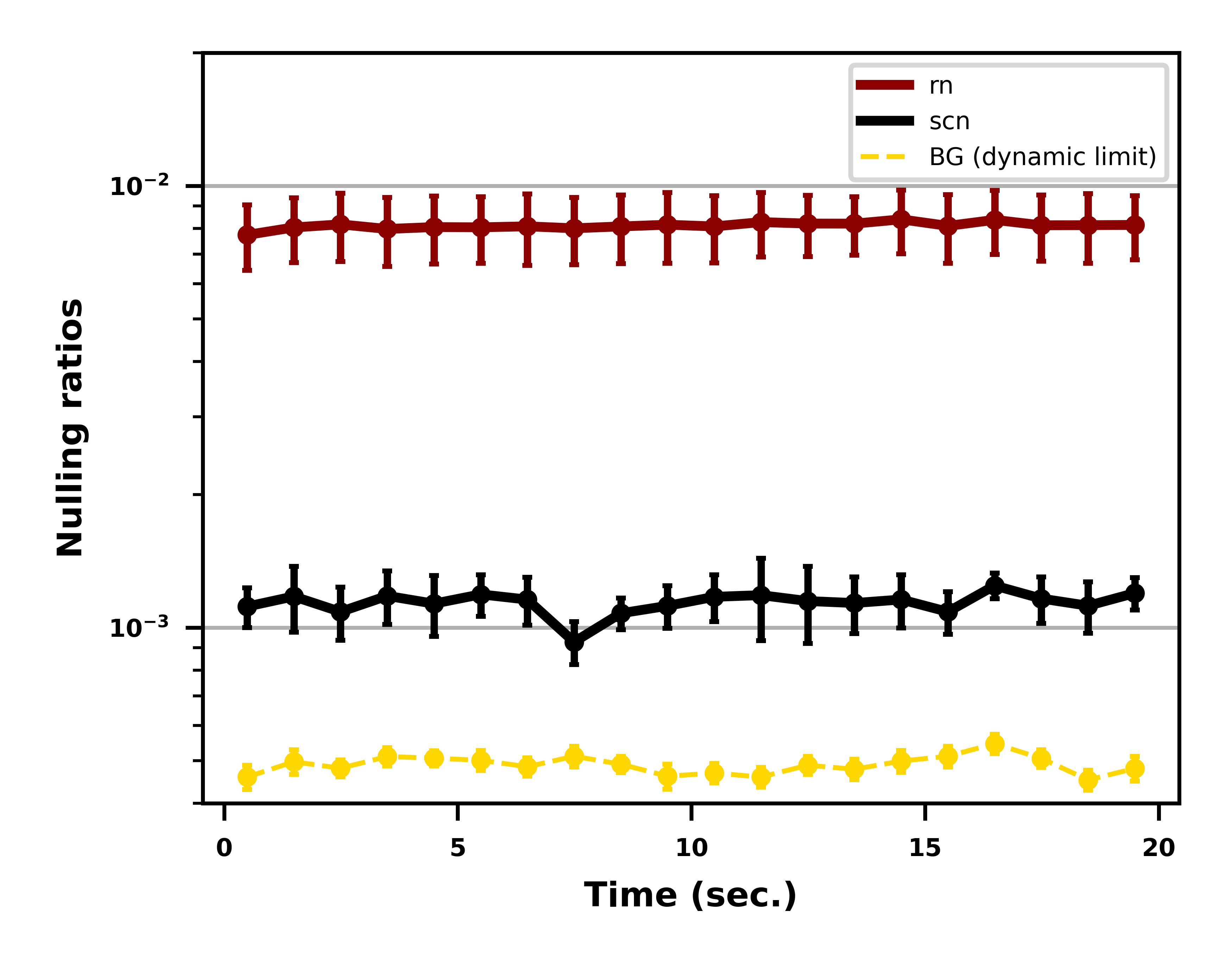}
    \caption{Self-calibrated (scn) and raw (rn) nulling ratio measured with the photonic 4T-nuller. The yellow dashed line shows the limit in the attainable dynamic range at room temperature set by the thermal background. Each point gives the mean value and the error on the mean estimated over 50 frames.
    }
    \label{fig:exp_rawnull_scnull}
\end{figure} 

\subsubsection{Null broadening using the double-Bracewell scheme}
\label{sec:results_star_transit}
\begin{figure}[t]
    \centering
    \includegraphics[width=1.0\columnwidth]{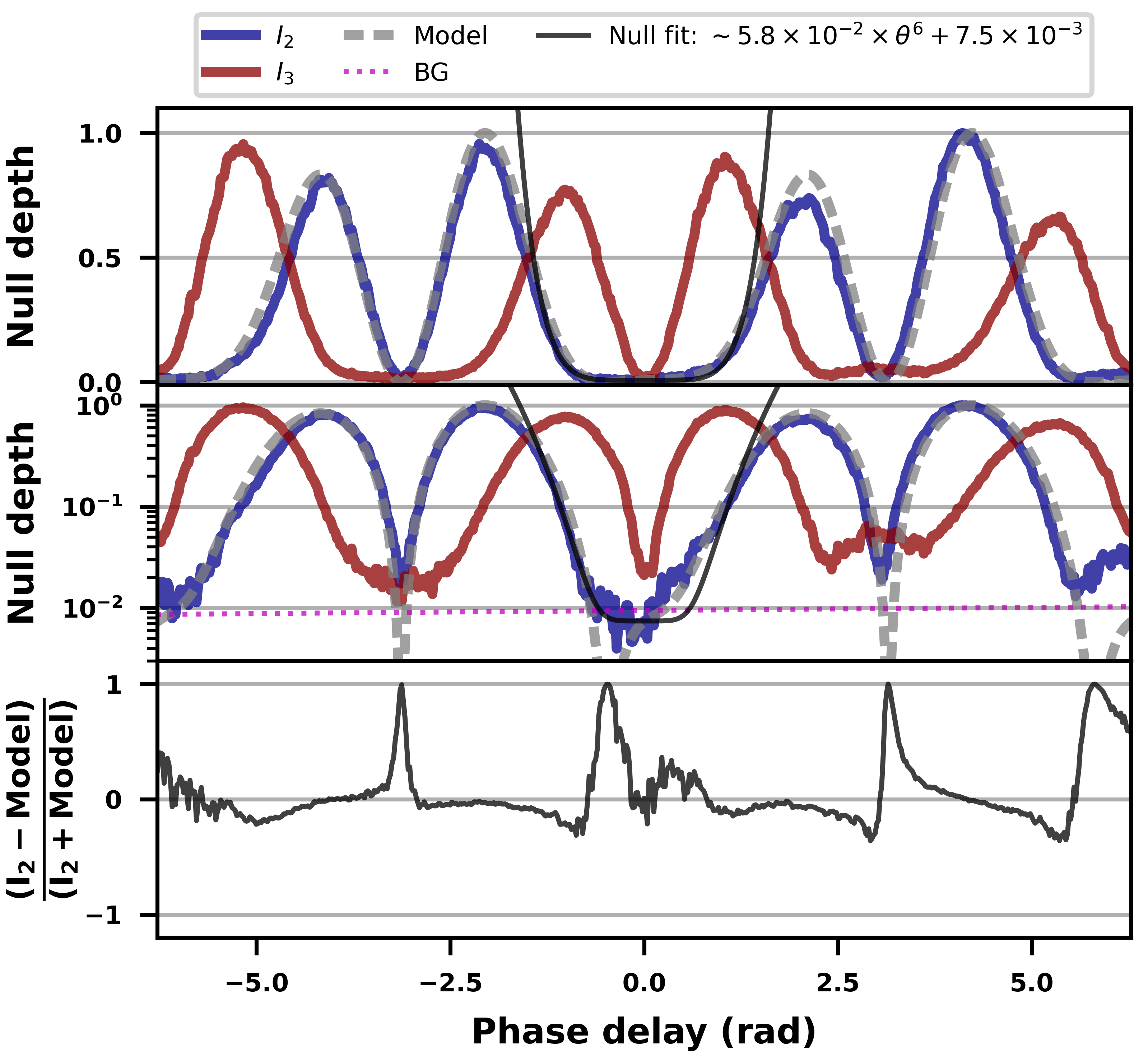}
    \caption{ 
    Transmission curves as a function of the phase delay of outputs $I_2$ and $I_3$, shown in blue and red, respectively, from the start transit simulation experiment, displayed in linear (top) and logarithmic scale (middle). The modified transmission model adapted from \cite{Angel1997} and fitted to the data is overlaid as a dashed gray line. The fit of the left part of the central broad null by a polynomial function is shown as a black continuous line. The background level is indicated with a dotted magenta line. The bottom plot shows the normalized fit residuals between the blue and dashed gray lines. 
    }
    \label{fig:NullTransmission_StarTransit}
\end{figure}
Following \cite{Errmann2015}, we aim at testing the use of our L$^{\prime}$ band ULI-fabricated beam combiner to implement the nulling scheme of \cite{Angel1997} that theoretically delivers a sixth-power dependence of the central null. The principle considers a linear interferometric array of four telescopes, with the two inner telescopes having twice the diameter of the two outer ones. The separation between the inner telescopes is 2$d$ while the outer telescopes are separated by a distance 4$d$. By moving our three delay lines of M2, M3, and M4 with the appropriate speeds, we are able to simulate a star transiting at the zenith of such an interferometer. \\
Using the layout of Fig.~\ref{fig:phasor}, the coupler DC1 forms a Bracewell nuller with the longest 4$d$ outer baseline, and the coupler DC2 forms a second Bracewell nuller with the shortest 2$d$ inner baseline. 
The two mirrors, M1 and M2, are steered so that the flux injected into inputs 1 and 2 is as close as possible to 1/4th of the flux injected into inputs 3 and 4 of the 4T-nuller. The initial relative phases must be $\pi$, $\pi/2$, 0 and $\pi/2$ for the inputs 1, 2, 3, and 4, respectively.  M1 being fixed, the speed of the delay lines of M2, M3, and M4 is set to 4$v$, 1$v$, and 3$v$, respectively, with the fundamental phase scan rate of $v = 0.44 \,$ rad/s. 
The camera acquires sequences of images of the 4T-nuller outputs (see Fig.~\ref{fig:4T-nuller-layout}) at a frame rate of 250\,Hz, and the intensity curves are extracted from the central pixels of outputs $I_2$ and $I_3$. The background is estimated for each frame by averaging the intensity of 16 pixels around the output pixels and subtracted from the frames to account for background variation due to laboratory temperature drift. The measurement is conducted in broadband (see Fig.~\ref{fig:bench}).\\ 
\noindent Fig.~\ref{fig:NullTransmission_StarTransit} (top) shows two fringe periods measured at the outputs $I_2$ and $I_3$ after scanning the three delay lines. The curves are smoothed using a six-point window. As expected, the interferograms from the two outputs are $\pi$ phase-shifted. 
The null depth at zero OPD is measured at $7.5 \pm 0.4 \times 10^{-3}$, but the measurement is limited by the background noise floor. $I_2$ (blue) shows the widened or broad null in broadband. $I2$ and $I3$ are both normalized against the peak flux of $I_2$. Note that the peak flux is only 480 camera counts due to the loss induced by decoupling the inputs of DC1 to reach the 1/4 th flux ratio between the outer and inner telescopes. \\ 
\noindent However, more information stands out in the logarithmic scale, as shown in Fig.~\ref{fig:NullTransmission_StarTransit} (middle). It reveals a slight deformation in the broad central null (blue), and the curve maxima do not all peak at the same value. We anticipate that this effect results from the limited synchronization accuracy of our delay lines, which extends to a latency of $\sim$100\,ms, as well as from their non-constant speed. The cause would be that the relative phase shifts between the beams deviate from the nominal values in the Angel\,\&\,Woolf scheme. To test this hypothesis, we fit a transmission model derived from the canonical model $T(\theta) = 4 \sin^{2}(\theta) \sin^{4}(\theta/2)$ of \cite{Angel1997}, in which additional small phase delays between the input beams are introduced. This model (gray dashes) fits the $I_2$ curve with phase delay parameters in the order of $\sim$0.03\,rad between the input beams. The level of agreement of this fitting is explored by looking at the normalized residual curve in Fig.~\ref{fig:NullTransmission_StarTransit} (bottom). We observe spikes in the residuals where the transmission dips reach the background level, which is expected since the background-free model predicts at these locations null levels that cannot be measured by our experiment. However, the variation among the transmission maxima and the deformation of the central null is adequately reproduced by the model. \\ 
\noindent In a second step of the analysis, we explored the $\theta$-dependence of the central broad null. We fitted the left portion of the central null with a polynomial function parameterized by $T(\theta) = a\theta^c + d$, as shown as a \textit{null fit} in the top and middle plots of Fig.~\ref{fig:NullTransmission_StarTransit}. 
The parameter $d$ = 7.5$\times$10$^{-3}$ corresponds to the vertical offset or the broad null depth at the zero OPD. From the fit results we find $a$\,=\,5.8$\pm$0.1$\times$10$^{-2}$ and $c$\,=\,6.0$\pm$0.1. The parameter $c$ is found to be in agreement with the theoretical model of \cite{Angel1997} for deep and broad central null, hence showing the ability of the photonic 4T-nuller to extinguish the light from the central star over a wide inner angle.
\subsubsection{Null degradation due to differential birefringence}
\label{sec:null_degradation_polarization}
To evaluate the effect of differential birefringence on the null, we employ a simple 2-beam interferometric model, where the differential phase delay between their polarization components, $s$ and $p$, is defined as $\phi_{\rm sp} = s - p$. The resulting polarization-dependent visibility degradation, $V_{\rm pol}$, can then be expressed as in Eq.~\eqref{eq:polarization} \citep{Traub1988}.
\begin{eqnarray}
V_{\rm pol} = |\cos(\phi_{\rm sp}/2)|
\label{eq:polarization}
\end{eqnarray}
\noindent For the 4T-nuller, the polarization-induced floor on the achievable raw null at $I_2$ can be estimated to $\sim$6.8$\times$10$^{-4}$ with the relation $rn = (1-V_{\rm pol})/(1+V_{\rm pol})$ based on Eq.~\eqref{eq:polarization}. Here, the polarization angle $\phi_{\rm sp}$=6$^\circ$ accounts for the average mismatch in polarization angle between input 2 and input 3 for $I_2$, as explained in Sect.~\ref{sec:results_polar_properties}. 
\noindent Fig.~\ref{fig:pola_angle} suggests that at 40$^{\circ}$ or 80$^{\circ}$ to 160$^{\circ}$, the degradation of visibility or null caused by differential birefringence is negligible $\sim$6.3$\times$10$^{-5}$, when $\phi_{\rm sp}\approx0$. However, at 18$^{\circ}$, the degradation due to differential birefringence increases to about $\sim$7.7×10$^{-3}$ as $\phi_{\rm sp}\approx18^{\circ}$. 
Fig.~\ref{fig:pola_contrast} also suggests that polarization contrast at 80$^{\circ}$ is highest when the linear polarization angle is horizontal relative to the vertical physical direction of the triplet (input facet).
However, the nulling performances described in Sect.~\ref{sec:results_extinction_ratios} and \ref{sec:results_star_transit} are conducted without controlling the polarization state (or without using any polarizer) of the injection beams. As a result, the polarization states of the input beams are unknown. Although the raw null degradation is estimated to be on the order of $\sim$4$\times$10$^{-3}$ for unpolarized (stellar) light entering the 4T-nuller, this represents a general case in which the integrated optics beam combiner induces circular polarization due to intrinsic birefringence effects. The detailed formalism is provided in Appendix~\ref{annex2.1}, and this result is further discussed in Sect.~\ref{sec:discussion}. Additionally, the minimum polarization contrast in Fig.~\ref{fig:pola_contrast} also indicates the level of phase retardation that causes birefringence, which is found to be $\Psi$$\sim$60$^{\circ}$, detailed in Appendix~\ref{annex2.2}.

\section{Discussion and conclusion} 
\label{sec:discussion}
We have reported the detailed characterization of the 4T-nuller that will be used for NOTT in the L$^{\prime}$ band. This result is, to our knowledge, the first measurement of a broadband L$^\prime$ deep null obtained with an integrated optics beam combiner. We obtained the following results:
\begin{itemize}
\item The 4T-nuller operates in single-mode regime between 3.65 and 3.85\,$\mu$m wavelength. It exhibits a flat, achromatic spectral response thanks to the asymmetric directional couplers. We demonstrated a 50/50 central directional coupler using 7.5\,mm interaction length, however, the side directional couplers show a less balanced 60/40 splitting ratio. Photometric taps tailored with achromatic 80/20 were achieved and will be integrated in a future version of the beam combiner.
\item The throughput is $\sim$37\% and will reach about 50\% once an anti-reflection coating is deposited. 
\item The emphasis was set to testing the high-contrast performance of the photonic combiner. We found that, in broadband operation and without polarization control, we could measure a raw null of 8.13$\pm$0.03$\times$10$^{-3}$ and a self-calibrated null of 1.14$\pm$0.01$\times$10$^{-3}$ over 20\,s. However, the raw null measurement does not reach the floor of the dynamic range of the testbed, with the nulled signal being larger than the background noise level. We suspect two types of limitations that we could however not disentangle. On the one hand, we suffer from a degradation of the null due to the broadband operation and the chromatic nature of our phase delay using free-space delay lines. We are also likely impacted by the residual imperfections in the 50/50 splitting ratio of the central combiner, and possibly to some degree of chromatic dispersion that is not characterized here. 
The latter effect will be corrected by the longitudinal dispersion compensator (LDC) of NOTT \citep{laugier_LDC_2024}.
On the other hand, the limited mechanical accuracy of our servo-motor delay lines prevented us from precisely fine-tuning the OPD to reach the deepest point of the null. This limitation will be mitigated during the final integration of the NOTT instrument thanks to Asgard's precise fringe tracker \citep{taras+asgard+fringe+tracker+2024}.
Interestingly, the implementation of the Angel\,\&\,Woolf scheme aiming at broadening the central null has shown that we could reach a raw null depth of $\sim$10$^{-2}$, which is limited by the floor of the thermal background. It is likely that in the conditions of a broadened null delivered by the Angel\,\&\,Wolf telescope configuration, the null depth is less sensitive to residual phase errors from the delay lines. 
\item The characterization of polarization properties indicates a low level of  differential birefringence. Exploiting the measured properties of the birefringence-induced polarization state of the output fields, we estimate a raw null degradation of 4$\times$10$^{-3}$ due to polarization mismatch, assuming the general case of two randomly circular polarized input beams. 
In this scenario, our estimate suggests that the reported raw null of null broadening measurement of $\sim$10$^{-2}$ is ultimately not limited by polarization mismatch.
However, because the true polarization state (i.e., the degree of elliptical polarization and its orientation) of our input beams on the bench was not independently characterized at the time of the experiment, we cannot yet estimate the precise impact of polarization mismatch on the raw null degradation reported here.
As the null degradation depends on the input polarization state, this effect can be classically mitigated by enforcing an input linear polarization along the angular direction where the differential birefringence is minimum. This could be done by splitting the incoming unpolarized light using a Wollaston prism (by treading off with a loss of sensitivity by 0.75 mag) or by modifying the input polarization states using a quarter-wave plate. 
A few caveats should be considered in the interpretation of these results: our measurement explored the properties of differential birefringence only for inputs 2, 3 and outputs $I_2$, $I_3$ of the 4T-nuller, which may not capture the full picture when operating with four input beams; improved measurements of the differential birefringence in the future should deliver a better sampling of the input polarization angle and give hints on the repeatability of such curves, which directly translates into a more robust level of uncertainty on the impact of polarization mismatches; our derivation of the impact of polarization mismatches only considers  monochromatic birefringence, whereas chromatic effects could play a role. In the context of a broadband characterization like ours, a previous characterization of the polarization state of the input beams used to measure the raw null will be required.\\
Given the importance of mitigating unwanted polarization effects in achieving high-contrast nulling interferometry, future directions should focus on developing more sophisticated strategies for polarization control and further improving the precision and repeatability of the ULI technology platform to minimize the level of differential birefringence. Finally, the post-coating polarization properties will have to be re-assessed.
\end{itemize}
The reported null ratios in this work must be regarded as pre-processing raw nulls and self-calibrated nulls. These results will undergo additional post-processing using nulling self-calibration numerically, which has been successfully implemented in PFN \citep{hanot_pnf_nuller_2011,Mennesson2011}
and LBTI \citep{defrere+BTI+2016} nullers’ data to achieve an interferometric nulling accuracy as low as a few times 10$^{-4}$. The extension of this technique to NOTT will use the GRIP \citep{martinod+grip+2025} and is currently under investigation. 
This aspect should be taken into account when comparing the reported experimental measurements to the requirement of 10$^{-5}$ reported in Table~\ref{tab:photonic_requirements}. 
It is important to highlight that the nulling ratios mainly depend on the 50/50 splitting behavior of the directional couplers within the 4T-nuller's interferometric section, apart from its differential birefringence properties and the imperfection of the delay lines. Consequently, the current 40/60 photometry/interferometry splitting of the Y-junctions, or a future upgrade to 20/80, will not compromise the 4T-nuller’s ability to achieve deeper nulls. In fact, operating the photonic chip under cryogenic conditions in the future will effectively reduce thermal background contributions present in the room-temperature laboratory setup, leading to improved nulling ratios.
\begin{acknowledgements} 
     A.S. acknowledges support from the Deutscher Akademischer Austauschdienst, Germany (DAAD Research Grants-57440920) and Macquarie University (iMQRES Scholarship-20191092) through a Cotutelle agreement, including partial support from the NASA Astrophysics Research and Analysis (APRA) program under grant number 80NSSC24K1560. This work was in part funded by the Australian Research Council Discovery Program under grant FT200100590. The work was performed in part at the OptoFab node of the Australian National Fabrication Facility (ANFF), utilizing Commonwealth as well as NSW state government funding. 
      KB acknowledges support from the Deutsche Forschungsgemeinschaft (DFG), grant number 506421303 ("NAIR-APREXIS").      
      DD, GG, MAM, and RL have received funding from the European Research Council (ERC) under the European Union's Horizon 2020 research and innovation program (grant agreement CoG-866070).
\end{acknowledgements}

   \bibliographystyle{aa} 
   \bibliography{reference} 

\appendix
\section{Impact of imbalanced directional couplers over the null}
\label{annex1}
\begin{figure}[b]
    \centering
    \includegraphics[width=0.47\columnwidth]{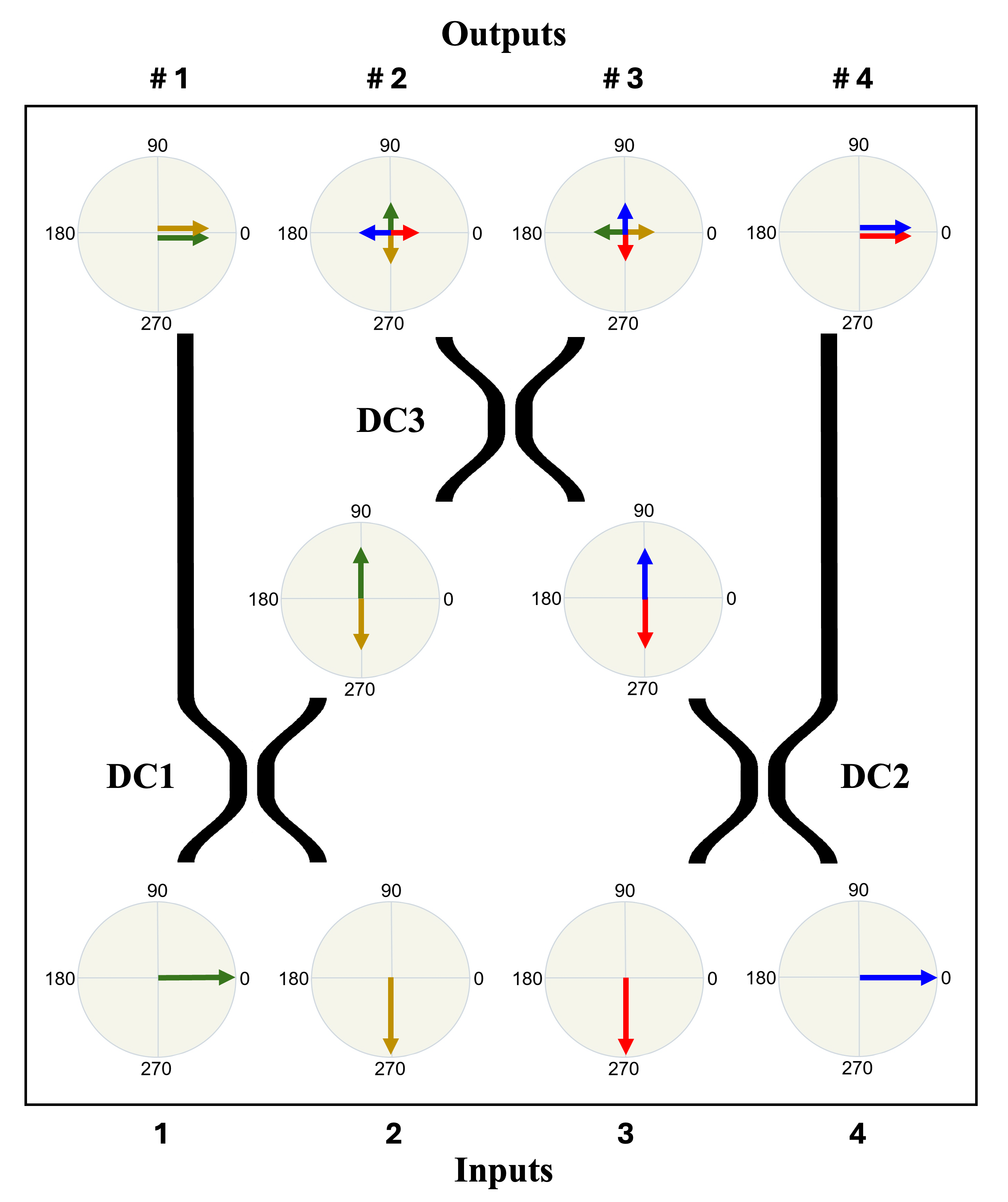} \hspace{0.2cm}
    \includegraphics[width=0.47\columnwidth]{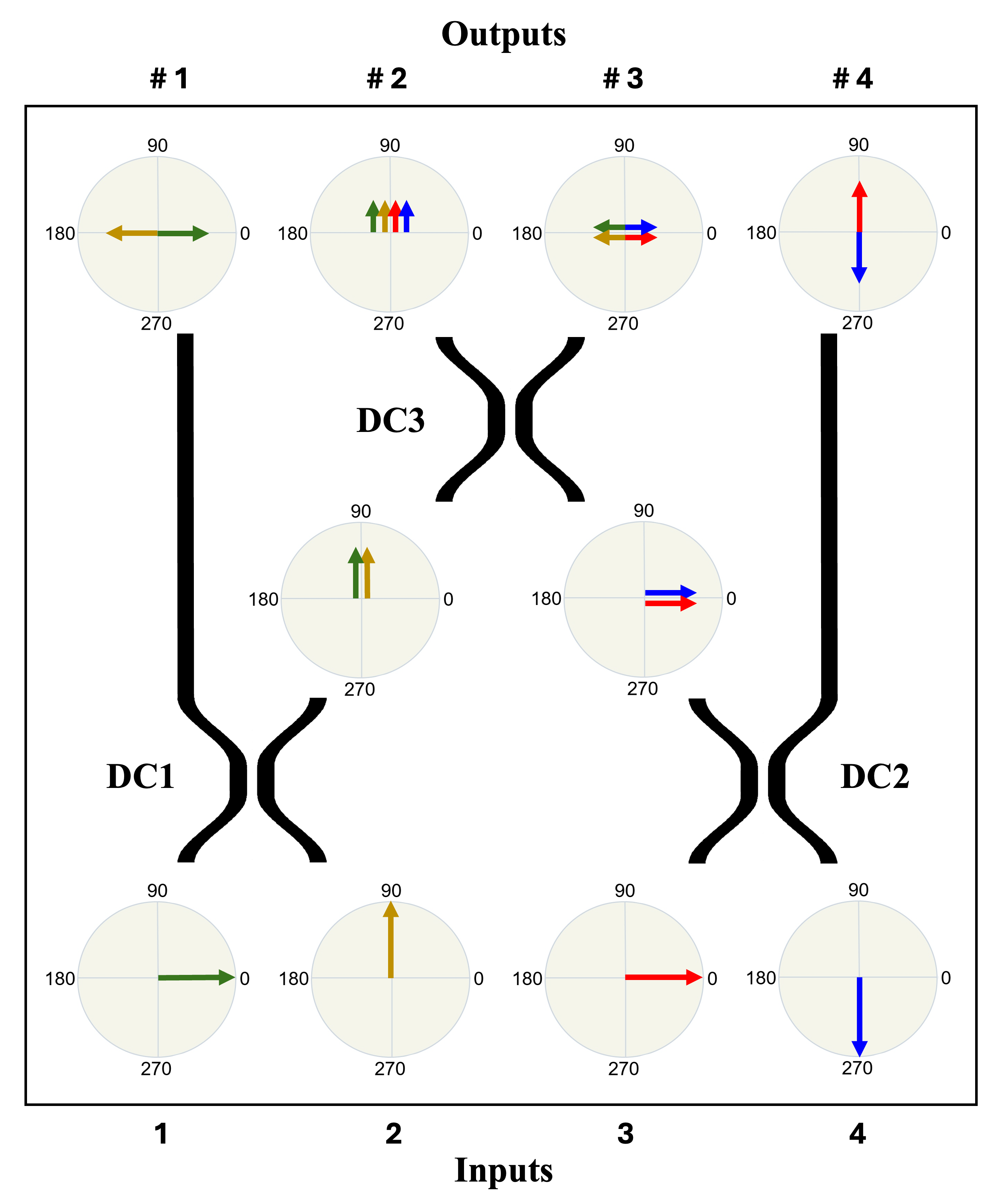}
    \caption{Phasor description of the input phase relationships in the destructive (left) and constructive (right) configurations of output \#2.}
    \label{ann:phasor}
\end{figure}

In this section, we describe how the deviation of DC1 and DC2 from a 50/50 splitting ratio can be compensated for in the derivation of the nulling ratio of Sect.~\ref{sec:results_extinction_ratios}.
At output \#2 of the photonic 4T-nuller where the planetary signal is detected\footnote{The same derivation applies if we consider output \#3} , the phase relationships to be considered are shown in the phasor of Fig.~\ref{ann:phasor}. These relationships would correspond to a destructive and constructive state for the star and the planet, respectively, in the nuller transmission map. In the following, we adopt for DC3 a splitting ratio of 50/50 and for DC1 and DC2 a splitting ration of 60(bar)/40(cross) as the result of our experimental characterization.\\
Let us denote $F_1$, $F_2$, $F_3$, and $F_4$ the flux coupled in the inputs 1, 2, 3, and 4, respectively, all with an arbitrary normalized coupling efficiency of 1. \\
\\
(1) In the constructive configuration, and assuming uniform propagation losses along the different optical paths followed by each beam, the flux at the output \#2 for the different input beams will be:
\begin{eqnarray}
O_2^1 &=& 0.4\cdot0.5\cdot F_1 \nonumber \\
O_2^2 &=& 0.6\cdot 0.5\cdot F_2 \nonumber \\
O_2^3 &=&0.6\cdot 0.5\cdot F_3 \nonumber \\
O_2^4 &=&0.4\cdot 0.5\cdot F_4 \nonumber
\end{eqnarray}
	
\noindent Because of the bench-related photometric imbalance among the four inputs described in Sect.~\ref{sec:setup}, the aforementioned four outputs will also experience a photometric imbalance. The weakest signal among these four outputs is the worst offender, and the photometric imbalance can be compensated for by decoupling the other three input beams until the four outputs $O_2^1$, $O_2^2$, $O_2^3$, and $O_2^4$ are all flux-equalized. Let us assume for simplicity that the worst offender is $O_2^1$. After carefully decoupling $F_2$, $F_3$ and $F_4$ to reach the photometric balance, the new values of the normalized coupling efficiency per beam are 1, ($F_1$/$F_2$)$\cdot$(0.4/0.6), ($F_1$/$F_3$)$\cdot$(0.4/0.6), ($F_1$/$F_4$). Clearly, the larger the initial photometric imbalance, the stronger the decoupling will need to be. In practice, the effective level of photometric imbalance intrinsic to the bench is at most about 15\% among the four input beams. Only a modest decoupling is required, and no noticeable scattered light is observed in the photonic chip, which could otherwise contaminate the measured output flux levels. \\
We now compute the total flux at output \#2 as the coherent addition of each beam amplitude according to 
\begin{eqnarray}
\lefteqn{	
I_2 = \left( \sqrt{0.4\cdot0.5\cdot F_1}+\sqrt{0.6\cdot0.5\cdot F_2\cdot(F_1/F_2)(0.4/0.6)} \right. } \nonumber \\ & &
\left. +\sqrt{0.6\cdot0.5\cdot F_3\cdot(F_1/F_3)(0.4/0.6)} \right. \nonumber \\ & &
\left. +\sqrt{0.4\cdot0.5\cdot F_4\cdot(F_1/F_4)}\, \right)^2 \nonumber
\end{eqnarray}
which simplifies into
\begin{eqnarray}
I_2 & = & 16\cdot 0.5\cdot 0.4\cdot F_1 \label{app:1}
\end{eqnarray}
(2) Following the same approach, the fluxes measured at the outputs \#1 and \#4 in the destructive configuration are given by
\begin{eqnarray}
I_1 &=& F_1\left(\sqrt{0.6}+\sqrt{\frac{0.4\cdot0.4}{0.6}}\right)^2 \label{app:2} \\
I_4 &=& F_1\left(\sqrt{\frac{0.4\cdot0.4}{0.6}}+\sqrt{0.6}\right)^2 \label{app:3}
\end{eqnarray}
The ratio of constructive fluxes measured in the two configurations is then given by
\begin{eqnarray}
\frac{I_2}{I_1+I_4} &=& \frac{16\cdot 0.5\cdot 0.4\cdot F_1}{2\,F_1\left(\sqrt{\frac{0.4\cdot0.4}{0.6}}+\sqrt{0.6}\right)^2} \nonumber
\end{eqnarray}
or $I_2$\,=\,0.96($I_1 + I_4$). \\
\\
In conclusion, in the presence of an imbalanced splitting ratio for DC1 and DC2, the simultaneous measurement of the constructive flux at the outputs \#1 and \#4 shall be compensated as explained in Sect.~\ref{sec:results_extinction_ratios}. \\
\\
(3) Finally, we can briefly generalize on a few aspects. If we assume a different flux worst offender than $O_1^2$, the expressions of Eq.~\ref{app:1},~\ref{app:2},~\ref{app:3} will be modified but the ratio $I_2$/($I_1+I_4$) remains unchanged. For instance, if we assume $O_3^2$ as the flux worst offender, we obtain
\begin{eqnarray}
\frac{I_2}{I_1+I_4} &=& \frac{16\cdot 0.5\cdot 0.6\cdot F_3}{2\,F_3\left(\sqrt{\frac{0.6\cdot0.6}{0.4}}+\sqrt{0.4}\right)^2} \nonumber
\end{eqnarray}
or $I_2$\,=\,0.96($I_1 + I_4$). \\
\\
For any identical, lossless bar/cross splitting ratio $\xi$/(1-$\xi$) of DC1 and DC2, the constructive flux measured from the outputs \#1 and \#4 can be related to the one measured at output \#2 through
\begin{eqnarray}
\frac{I_2}{I_1+I_4} &=& \frac{16\cdot 0.5\cdot \xi}{2\,\left(\sqrt{\frac{\xi^2}{(1-\xi)}}+\sqrt{1-\xi)}\right)^2} \nonumber
\end{eqnarray}
which shows, as expected, that $\xi$=0.5 is the ideal, balanced case.

\section{Beam combiner birefringence}\label{annex2}
\subsection{Null degradation due to differential birefringence}\label{annex2.1}

\begin{figure}[h]
    \centering
    \includegraphics[width=\columnwidth]{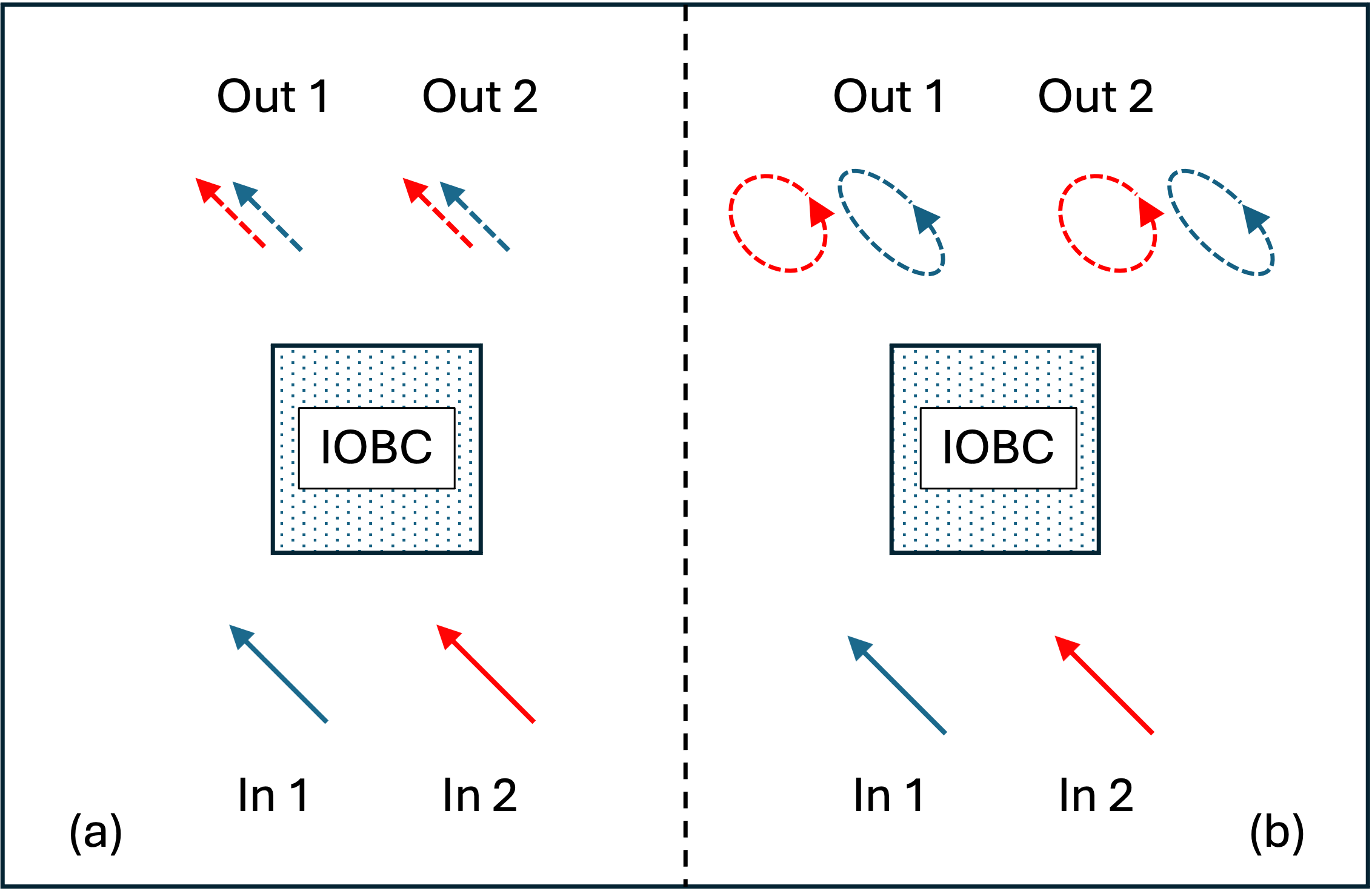}
    \caption{(a) In an ideal scenario with no birefringence in the integrated optics beam combiner (IOBC), any input linear polarization state remains preserved, as shown with blue and red arrows. The electric fields combined interferometrically at the output, either Out 1 or Out 2, are identical, so no polarization-induced null degradation occurs. (b) In practice, when birefringence is present in the IOBC, input linear polarization transforms into different elliptical polarization states, as shown in Out 1 and Out 2, if it is not aligned with the waveguide or optical path's fast or slow axis, which can cause potential null degradation.
}
    \label{ann:biref}
\end{figure}

The impact of differential birefringence on polarization mismatch and, consequently, on null degradation is illustrated in Fig.~\ref{ann:biref}, which compares the ideal and practical cases for a monochromatic signal using a $2 \times 2$ integrated optics beam combiner.\\
For a given output, Out 1 of Fig.~\ref{ann:biref}(b), the components of the elliptically polarized output fields $\boldsymbol{E_1}$ and $\boldsymbol{E_2}$ can be written as
\begin{eqnarray}
\lefteqn{
E_{\rm 1x}= \left. \sqrt{1+\alpha_{\rm 1}}\cos(\omega t)\cos(\theta_{\rm 1}) \right. } \nonumber \\ & &
\hspace{2cm}\left. +\sqrt{1-\alpha_{\rm 1}}\sin(\omega t)\cos(\frac{\pi}{2}+\theta_{\rm 1}) \right. \nonumber \\ 
\lefteqn{
E_{\rm 1y}= \left. \sqrt{1+\alpha_{\rm 1}}\cos(\omega t)\sin(\theta_{\rm 1}) \right. } \nonumber \\ & &
\hspace{2cm}\left. +\sqrt{1-\alpha_{\rm 1}}\sin(\omega t)\sin(\frac{\pi}{2}+\theta_{\rm 1}) \right. \nonumber \\ 
\lefteqn{
E_{\rm 2x}= \left. \sqrt{1+\alpha_{\rm 2}}\cos(\omega t+\varphi)\cos(\theta_{\rm 2}) \right. } \nonumber \\ & &
\hspace{2cm}\left. +\sqrt{1-\alpha_{\rm 2}}\sin(\omega t+\varphi)\cos(\frac{\pi}{2}+\theta_{\rm 2}) \right. \nonumber \\ 
\lefteqn{
E_{\rm 2y}= \left. \sqrt{1+\alpha_{\rm 2}}\cos(\omega t+\varphi)\sin(\theta_{\rm 2}) \right. } \nonumber \\ & &
\hspace{2cm}\left. +\sqrt{1-\alpha_{\rm 2}}\sin(\omega t+\varphi)\sin(\frac{\pi}{2}+\theta_{\rm 2}) \right. \nonumber
\end{eqnarray}
where $\alpha_{\rm 1,2}$ is the polarization contrast (i.e., $P_c$ in Eq.~\ref{eq:contrast}) and $\theta_{\rm 1,2}$ the angle between the horizontal axis and major axis of the ellipse. The values of $\alpha_{\rm 1,2}$ and $\theta_{\rm 1,2}$ are measured in Fig.~\ref{fig:pola_contrast} and \ref{fig:pola_angle}. The variable $\omega$ is the frequency and $\varphi$ is the variable phase delay induced by the delay line. From these relations, few special cases can be noticed. For $\alpha_{\rm 1}$=$\alpha_{\rm 2}$=1 and $\theta_{\rm 1}$=$\theta_{\rm 2}$=0, $\boldsymbol{E_1}$ and $\boldsymbol{E_2}$ have a linear horizontal polarization. If $\theta_{\rm 1}$=$\theta_{\rm 2}$=$\pi$/2, $\boldsymbol{E_1}$ and $\boldsymbol{E_2}$ have a linear vertical polarization. For $\alpha_{\rm 1}$=$\alpha_{\rm 2}$=0, $\boldsymbol{E_1}$ and $\boldsymbol{E_2}$ are circularly polarized.\\
Using this set of equations and the respective values of $\alpha$ and $\theta$ from Fig.~\ref{fig:pola_contrast} and \ref{fig:pola_angle}, it is possible to compute, for any angular direction of the input linear polarization, the total time-averaged intensity resulting from the interferences between $\boldsymbol{E_1}$ and $\boldsymbol{E_2}$. Afterwards, the interferometric contrast $v$ is computed by varying the phase delay $\varphi$. The corresponding raw null due to polarization mismatch is obtained as $rn$=(1-$v$)/(1+$v$). In this way, we can estimate the raw null degradation resulting from differential birefringence in cases where the electrical fields at inputs 1 and 2 exhibit either controlled linear polarization or random circular polarization. As Fig.~\ref{fig:pola_contrast} and \ref{fig:pola_angle} only report discrete values in steps of 20$^{\circ}$ of the input polarization angle, we retrieve all other values via spline interpolation. Note that possible chromatic effects of the waveguide birefringence are not considered here.

\subsection{Phase retardation between the fast and slow axis of the birefringent waveguide}
\label{annex2.2}
The characterization of the polarization contrast can serve as a proxy of the birefringence-induced phase retardation between the fast and slow axis. The minimum polarization contrast measured in Fig.~\ref{fig:pola_contrast} is related to the phase retardation $\Psi$ as \citep{Benoit2021}
\begin{eqnarray}
    P_{c,\rm min}&\approx&|\cos(\Psi)| \nonumber
\end{eqnarray}
For instance, if we measure a polarization contrast $P_{c,\rm min}$=1 for any angle of the input linear polarization, the linear polarization state is always preserved between the input and the output. This corresponds to the case where the phase retardation $\Psi$=0$^{\circ}$ (or $\Psi$=180$^{\circ}$), i.e., there is no birefringence present. In contrast, if $P_{c,\rm min}$=0, the input linear polarization can be transformed into an output circular polarization. In this case, a phase retardation $\Psi$=90$^{\circ}$ is required.

\end{document}